\documentclass[prx,%
 reprint,
 amsmath,amssymb,
 aps,prx
]{revtex4-2}
\usepackage[english]{babel}

\usepackage{amsmath}
\usepackage{bm}
\usepackage{amsfonts}
\usepackage[cal=cm]{mathalpha}
\usepackage{nicematrix}
\usepackage{graphicx}
\usepackage{tcolorbox}
\usepackage[colorlinks=true, allcolors=blue]{hyperref}
\usepackage{amsthm}
\usepackage{algorithm}
\usepackage{algpseudocode}
\usepackage{epstopdf}
\usepackage{booktabs}
\usepackage{lineno}

\begin{document}

\title{Mesoscale community organization governs epidemic onset and spread in metapopulations}

\author{Haoyang Qian} 
\author{Malbor Asllani} 
 \affiliation{Department of Mathematics, Florida State University,
1017 Academic Way, Tallahassee, FL 32306, United States of America}

\begin{abstract}
Understanding how internal community structure shapes the course of epidemics remains a fundamental challenge in modeling real-world populations. Standard metapopulation models often assume uniform mixing within communities, overlooking how internal heterogeneity affects global outcomes. Here, we develop a general framework for epidemic spreading in hierarchically structured metapopulations, where individuals interact locally within dense communities and move across a broader network. We show that transmission dynamics are governed by the mesoscale organization of these communities: highly connected groups accelerate and amplify outbreaks, while less connected ones dampen spread. Through a combination of mean-field theory, spectral analysis, and stability methods, we reveal a direct link between internal connectivity and the emergence of uneven, spatially structured epidemic patterns. We further validate these predictions using real-world data, where social contact networks capture the local scale of transmission while spatial transport networks govern global connectivity, confirming the robustness of our framework across scales. These results demonstrate how community structure fundamentally governs the shape of epidemics in complex, networked populations, offering new insights into vulnerability, containment, and epidemic control.
\end{abstract}

\maketitle

\section{Introduction}
\label{sec:intro}

Recent health crises have underscored the need for a better understanding of how behavioral responses and population heterogeneity influence disease transmission dynamics~\cite{vespignani2020modelling, malvy2019ebola}. Preventive behaviors and social compliance can amplify or suppress outbreaks in ways that standard models often fail to capture~\cite{morsky2019evolution}. Key topological features—such as degree heterogeneity, clustering, and community organization—can drastically alter epidemic thresholds and spreading speed~\cite{spreading_book, epidemics_review}, informing the design of targeted interventions including immunization strategies~\cite{pastor2002immunization}, behavioral feedback mechanisms~\cite{granell2013dynamical}, and mobility-aware containment policies~\cite{siebert2022nonlinear}.

Network-based approaches have proven essential for representing individual-level interactions, where nodes and edges encode population structure and contact patterns~\cite{newman_book, estrada_book}. Within this framework, two complementary scales of modeling have emerged. Contact networks focus on local interactions within social or physical spaces—such as households, schools, or workplaces—highlighting how heterogeneous connectivity (e.g., scale-free structures) can eliminate epidemic thresholds and drive widespread contagion~\cite{pastor2001epidemic_1, pastor2001epidemic_2}. However, when disease spreads across spatially separated communities, local contact patterns become insufficient. Instead, mobility networks model inter-community movement, where nodes represent population centers and edges capture travel or commuting routes~\cite{colizza2007reaction, balcan2009multiscale, wang2022epidemic}. These models often adopt a reaction-diffusion approach, where transmission occurs locally while individuals migrate between patches~\cite{colizza2006role, brockmann2013hidden}. Accurately capturing epidemic dynamics in structured populations requires integrating both local interaction structures and global mobility pathways, as their interplay fundamentally shapes the spatial and temporal progression of disease.

Despite increasing attention to epidemic dynamics in structured populations, most existing models tend to emphasize either large-scale mobility across a diffusion network or the internal structure of local contact networks—seldom integrating both perspectives within a unified framework~\cite{phase_struct}. Models often treat patches as well-mixed compartments~\cite{murray, caswell}, with infection rates inferred empirically rather than derived from network topology. However, recent studies underscore the importance of incorporating community-level structure to understand how local interactions influence outbreak trajectories. This includes work on group-structured populations linking membership patterns to disease severity~\cite{patwardhan2023epidemic}, and on multiplex metapopulations where overlapping mobility layers reshape transmission pathways~\cite{soriano2018spreading}. Other approaches highlight how the coupling of local dynamics with long-range movement gives rise to complex spreading patterns~\cite{brockmann2013hidden, allard2017geometric, NJP}. While these contributions represent important steps toward multiscale modeling, they often fall short of explicitly capturing the heterogeneity of local contact networks alongside global mobility—leaving a conceptual and methodological gap in the representation of structured populations.

In this work, we develop a comprehensive mathematical framework that integrates local interaction dynamics with global mobility by embedding contact networks within a metapopulation structure. This dual-scale formalism captures both intra- and inter-community heterogeneity, offering a more realistic representation of structured populations. Our framework derives infection rates directly from the underlying network structure, moving beyond standard models that assume homogeneous infection rates. To retain analytical tractability while capturing essential structural heterogeneity, we employ a mean-field approximation that yields an effective, structure-dependent contagion rate for each metanode.

A central contribution of our formalism lies in demonstrating, via spectral perturbation theory~\cite{franklin2012matrix, horn2012matrix}, that communities with higher-than-average internal connectivity act as dominant drivers of epidemic propagation across the metapopulation. This coarse-grained insight is further refined by leveraging the localization properties of Laplacian eigenvectors, which tend to concentrate activity on a subset of nodes in large, disordered networks~\cite{mcgraw, nakao_loc, pastor2016distinct}. We use this localization to construct a reduction method that decouples the influence of localized modes from the rest of the system, revealing how dense or highly connected metanodes disproportionately shape the epidemic’s global footprint. This spectral perspective establishes a direct link between network structure and spatial contagion patterns, providing predictive insight into the role of mesoscale organization. While our analysis focuses on epidemic dynamics, the reduction method can be broadly applicable to other structured population models, offering a versatile tool for analyzing high-dimensional dynamics in complex systems.



\section{Individual-based Modeling of Spreading dynamics in Hierarchically Structured Metapopulations}
\label{sec:II}

\begin{figure}[t!]
 \centering
 \includegraphics[width=\columnwidth]{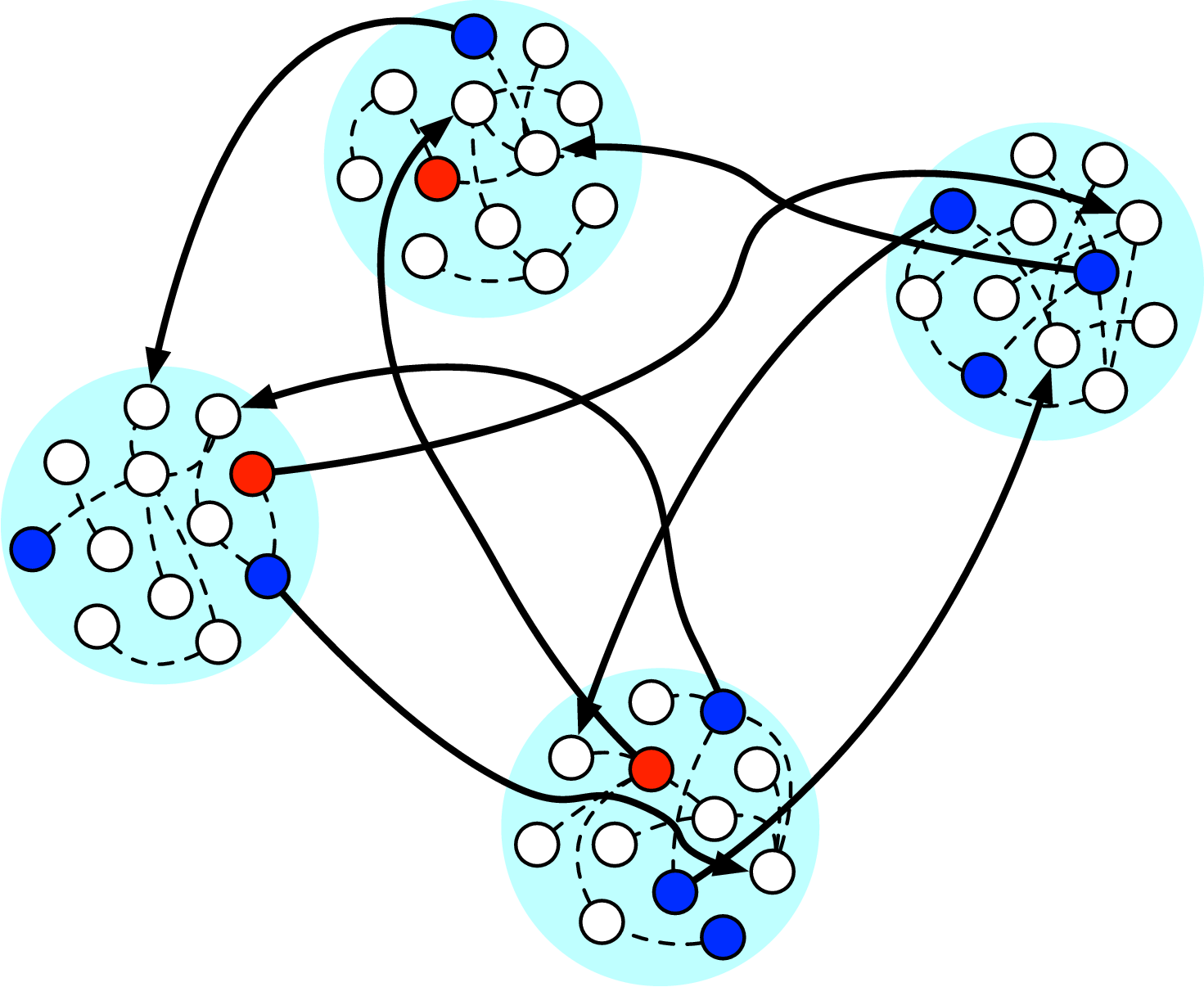}
\caption{Schematic of the spreading process on a hierarchically structured network. Individuals—either susceptible \( S \) (blue) or infected \( I \) (red)—can interact (dashed links) when co-located within the same spatial patch, represented by a metanode (light blue). Migration between metanodes (solid links) occurs via assigned positions within the destination metanode, ensuring individuals are associated with a single metanode at any given time.}
 \label{fig:metaplex}
 \end{figure}

In this section, we introduce a model of a hierarchically structured metapopulation network, where local contact networks are embedded within a global diffusion structure—allowing the integration of individual-level interactions with large-scale mobility. This formulation generalizes classical mean-field approaches by capturing the multiscale organization of real-world epidemics, while remaining analytically tractable. Although such models could be derived from microscopic dynamics via master equations~\cite{van1992stochastic, gardiner}, we instead adopt a mean-field framework that naturally couples metapopulation diffusion~\cite{diffusion_barrat} with local contact processes~\cite{epidemics_review}. In this setting, agents move between spatial patches (metanodes) via diffusion, while interacting through fixed contact networks within each patch—representing social contexts such as schools, workplaces, or households. Disease transmission occurs only within local networks, conditional on co-location and connectivity. As illustrated in Fig.~\ref{fig:metaplex}, each individual occupies a uniquely defined position in every metanode, and movements occur only when a vacant position is available. It highlights both the inter-node displacement (black arrows) and the internal contact links (dashed lines), distinguishing mobility from local transmission.

While other contagion processes can be readily incorporated, here we focus on a Susceptible-Infected-Susceptible (SIS) model that captures both individual-level interactions and group-level mobility. The dynamics are governed by the \textit{Individual-Based Mean-Field} (IBMF) formulation:
\begin{equation}
\label{eq:IBMF}
\begin{aligned}
    \dot{S}_{i,\mu} &= -\beta \sum_{j=1}^N A^{(\mu)}_{ij} S_{i,\mu} I_{j,\mu} + \gamma I_{i,\mu} + D_S \sum_{\nu=1}^{\Omega} \mathcal{L}_{\mu \nu} S_{i,\nu}, \\
    \dot{I}_{i,\mu} &= \beta \sum_{j=1}^N A^{(\mu)}_{ij} S_{i,\mu} I_{j,\mu} - \gamma I_{i,\mu} + D_I \sum_{\nu=1}^{\Omega} \mathcal{L}_{\mu \nu} I_{i,\nu}, \quad \forall i, \mu
\end{aligned}
\end{equation}
where \( S_{i,\mu} \) and \( I_{i,\mu} \) denote the probabilities that individual \( i \) in metanode \( \mu \) is susceptible or infected, respectively. Here, \( i \) indexes individuals and \( \mu \) denotes metanodes, each of which accommodates at most \( N \) individuals simultaneously. The contagion process within each metanode is governed by the adjacency matrix \( \textbf{A}^{(\mu)} \), which defines local contact networks: \( A^{(\mu)}_{ij} = 1 \) indicates that individuals \( i \) and \( j \) are in contact while co-located in metanode \( \mu \), leading to transmission at rate \( \beta \), whereas infected individuals recover independently at rate \( \gamma \). The diffusion terms describe the movement of individuals across metanodes via the graph Laplacian $\boldsymbol{\mathcal{L}}$ with entries \( \mathcal{L}_{\mu \nu} = \mathcal{A}_{\mu \nu} - k_\mu \delta_{\mu \nu} \), where \( \mathcal{A}_{\mu \nu} \) defines the metanode connectivity matrix and \( k_\mu \) is the degree of metanode \( \mu \). These terms ensure that mobility follows the topology of the metapopulation network. The IBMF formulation provides a flexible platform for studying contagion, which can be adapted to various mean-field approximations—including individual-based~\cite{kiss_mathematics_2017}, degree-based~\cite{pastor2001epidemic_1, pastor2001epidemic_2}, pairwise~\cite{gleeson2013binary}, or higher-order~\cite{battiston2021physics, ferraz2024contagion} models—depending on the complexity and resolution required. For simplicity, we consider undirected networks at both mobility and contact levels, though the framework naturally extends to directed cases often relevant in real-world settings~\cite{asllani2018topological, NN}.

In classical metapopulation models, the local interaction structure is typically omitted and replaced by an effective contagion rate, empirically estimated to account for contact heterogeneity~\cite{spreading_book}. By contrast, our two-level framework introduced in Eq.~\eqref{eq:IBMF} enables a more principled derivation of transmission rates, grounded in the explicit topology of local interactions. A key distinction from standard contact-network models is that the probability of an individual \( i \) being in a given state at metanode \( \mu \) does not sum to unity, i.e., \( S_{i,\mu}(t) + I_{i,\mu}(t) \neq 1 \), since the individual may not be present at \( \mu \) at time \( t \). However, because the system is closed, each individual must be in some state somewhere in the network, ensuring the normalization condition \( \sum_{\mu=1}^{\Omega} \left(S_{i,\mu}(t) + I_{i,\mu}(t)\right) = 1 \) holds for all \( t \). In what follows, we focus on estimating the local contagion rate, laying the foundation for a mathematically rigorous analysis of the spreading dynamics.

\begin{figure*}
    \centering
    \includegraphics[width=\textwidth]{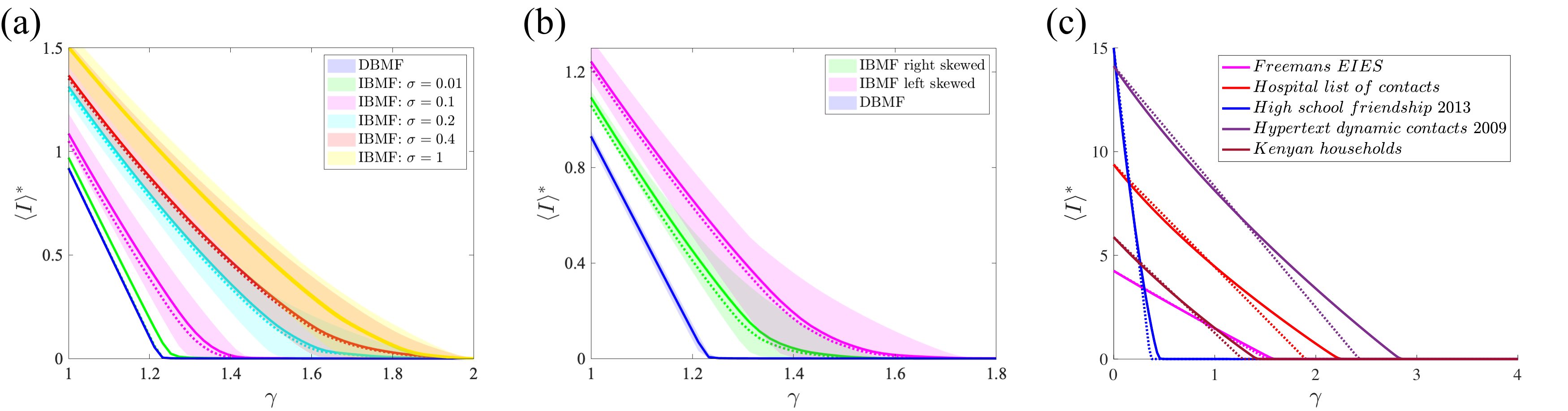}
\caption{The mean equilibrium state, \(\langle x \rangle^*\), is plotted against the recovery rate, \(\gamma\) for different networks parameters realizations. In the metapopulation network, both the transport and contact networks are Erdős–Rényi graphs with wiring probabilities \( q \) (between metanodes) and \( p \) (within each contact network), respectively. For both panels, each scenario corresponding to a different \( p \) value list is simulated 10 times using both the IBMF and DBMF\(_{\mu}\) frameworks. In each simulation, a distinct distribution of \(\langle k \rangle_{\mu}\) is generated. The shaded regions in the IBMF results represent the range of observed values across the 10 simulations, with the solid line within the shaded region indicating the mean outcome. The dashed line corresponds to the average result obtained from DBMF\(_{\mu}\) simulations. The shaded region is constructed by setting its boundaries using the minimum and maximum values of $\langle k \rangle_{\mu}$ obtained from the simulations. The solid line within this region represents the mean value obtained from the IBMF simulations.
(a) The distribution of \(\langle k \rangle_{\mu}\) follows a normal distribution with a mean of 0.5 and varying variance $\sigma$. 
(b) The distribution of \( p \) is systematically varied to include both left-skewed and right-skewed forms while maintaining a similar mean. Specifically, a gamma distribution (\(\alpha = 1, \theta=2\)) is utilized to generate left-skewed \( p \), and the right-skewed \( p \) is obtained by applying a reflection transformation to the left-skewed distribution. (c) \emph{Empirical social networks}: we apply the same protocol to fixed real social graphs, plotting $\langle x \rangle^{*}$ versus $\gamma$ with IBMF (solid) and DBMF$_{\mu}$ (dashed); dataset definitions and analysis are detailed in the Supplemental Information.
 The set of parameters for all panels are $\Omega=10, N=50, q=0.5, \beta=0.5, D_S=0.2, D_I=0.1$.}
     \label{fig:Fig1}
\end{figure*}


\section{Degree-Based Mean-Field Approximation and Numerical Results}
\label{sec:III}

To reduce the dimensionality of the IBMF model and obtain a tractable coarse-grained description, we derive an effective system for the total probabilities \( S_\mu = \sum_{i=1}^{N} S_{i,\mu} \) and \( I_\mu = \sum_{i=1}^{N} I_{i,\mu} \) of susceptibles and infecteds in each metanode \( \mu \). A detailed derivation is provided in the Appendix \ref{sec:DBMF_App}; here we briefly sketch the main ideas. Individual-level variables are decomposed into a metanode-based mean-field part and a deviation term: \( S_{i,\mu} = \langle S_\mu \rangle + \delta S_{i,\mu} \) and \( I_{i,\mu} = \langle I_\mu \rangle + \delta I_{i,\mu} \). To estimate the contribution of the contact network, we apply a standard degree-based approximation, assuming \( S_{i,\mu} \sim k_i \) and \( I_{i,\mu} \sim k_i \), consistent with widely used methods in network epidemiology~\cite{pastor2001epidemic_1, pastor2001epidemic_2}. This step requires that the degree distribution \( P(k) \) within each metanode be sufficiently narrow and symmetric. Under these assumptions the resulting infection term scales as \[ \sum_{i,j} A^{(\mu)}_{ij} S_{i,\mu} I_{j,\mu} \approx  N \langle k \rangle_\mu \langle S \rangle_\mu \langle I \rangle_\mu\,, \] where, from now on, we omit the summation limits for simplicity. This yields an effective contagion rate \( \tilde{\beta}_\mu = \beta \langle k \rangle_\mu / N \) that captures the local structural heterogeneity. The resulting reduced system reads:
\begin{equation}
\label{eq:DBMF_mu}
\begin{aligned}
\dot{S}_{\mu} &= -\tilde{\beta}_{\mu} S_{\mu} I_{\mu} + \gamma I_{\mu} + D_S \sum_{\nu} \mathcal{L}_{\mu \nu} S_{\nu}, \\
\dot{I}_{\mu} &= \tilde{\beta}_{\mu} S_{\mu} I_{\mu} - \gamma I_{\mu} + D_I \sum_{\nu} \mathcal{L}_{\mu \nu} I_{\nu}, \quad &\forall \mu.
\end{aligned}
\end{equation}
We refer to Eqs.~\eqref{eq:DBMF_mu} as the \emph{Degree-Based Mean-Field} (DBMF\(_\mu\)) system, as it relies on the standard degree-based approximation and uses the mean degree \( \langle k \rangle_\mu \) to represent each community’s contact structure.

To establish a clear baseline for comparison with the heterogeneous case, we adopt a homogenized formulation of the DBMF\(_\mu\) system by averaging over the structural heterogeneity of metanodes. Specifically, we replace the metanode-specific effective transmission rates \( \tilde{\beta}_\mu \) with a uniform quantity \( \tilde{\beta} = \beta \langle \langle k \rangle_\mu \rangle / N \), where \( \langle \langle k \rangle_\mu \rangle = \sum_{\mu=1}^\Omega \langle k \rangle_\mu / \Omega \) denotes the average of the mean degrees across all metanodes. This simplification, referred to as the \emph{DBMF} model, preserves the essential interplay between diffusion and contagion dynamics while enabling a mathematically transparent representation suitable for the stability analysis—otherwise hindered by the intractability of heterogeneous transmission rates~\cite{sun2009master, nazerian2023synchronization}.

Figure~\ref{fig:Fig1} reports the mean equilibrium infection level as a function of the recovery rate, based on multiple simulations across the IBMF, DBMF, and DBMF\(_\mu\) frameworks. This comparison highlights the role of community-level degree heterogeneity in shaping disease dynamics. When \( \langle k \rangle_\mu \) varies across metanodes, the infection persists more easily than in the homogeneous DBMF case. In panel~(a), each realization corresponds to a distinct distribution of \( \langle k \rangle_\mu \); shaded regions represent the variability of IBMF outcomes, while the solid line shows their mean. The dashed line marks the corresponding DBMF\(_\mu\) prediction, which closely tracks the IBMF average—demonstrating that DBMF\(_\mu\) captures the essential dynamics despite underlying heterogeneity.

Panel~(a) reveals a key finding: increasing community heterogeneity sustains higher infection levels, even for large recovery rates. When internal connection probabilities follow a normal distribution, increasing variance leads to a growing discrepancy between IBMF and DBMF predictions—underscoring the importance of mesoscale structural variability. This mechanism is central to our analysis and will be explored in the following section. Panel~(b) further shows that left-skewed distributions yield higher prevalence than right-skewed ones with similar means, as the bulk of high-degree communities concentrated on the right side of the distribution enhances transmission. These hubs act as accelerators of contagion, while the sparse low-degree communities in the left tail contribute minimally to the spread. In panel (c), we test the degree–based closure DBMF$_\mu$ against the individual–based benchmark (IBMF) on {empirical social networks}, using fixed, time–aggregated face-to-face graphs from a conference (HT09), a high school, a workplace, and a hospital ward (datasets listed in the inset; details in the SM). Across this recovery-rate sweep $\gamma$, DBMF$_\mu$ closely reproduces the IBMF predictions for the mean equilibrium state $\langle x\rangle^{*}$, validating the degree-based approximation as a reliable surrogate on real interaction topologies.


%
%
%


\section{Perturbation analysis for the heterogeneous community densities}
\label{sec:V}

As anticipated, analyzing a general spatially extended reaction-diffusion system becomes analytically intractable when heterogeneous parameters are introduced, obscuring the mechanisms by which denser communities promote disease spread and shape other key epidemic features. To overcome this, we adopt a weak formulation where contact network densities deviate slightly from their mean, enabling a perturbative analysis of the eigenvalues and eigenvectors of the Jacobian matrix associated with the DBMF\(_\mu\) model. As shown in the following sections, the perturbed Jacobian’s eigenvectors and eigenvalues are crucial for explaining the contrasting roles of dense and sparse communities during both early and later stages of the epidemic. For completeness, a brief overview of spectral perturbation theory, including eigenvector corrections, is provided in Appendix \ref{sec:perturb}.

We begin by examining the structure of the Jacobian \( \boldsymbol{\mathcal{\tilde{J}}} \), a \( 2\Omega \times 2\Omega \) block matrix that will later represent the first-order term in a perturbative expansion around the homogeneous case
\[
\boldsymbol{\mathcal{\tilde{J}}} = \boldsymbol{\mathcal{{D}}} + \boldsymbol{\mathcal{J}},
\]
where  
\[
\boldsymbol{\mathcal{{D}}} = 
\begin{pmatrix}
D_S \boldsymbol{\mathcal{L}} & \mathbf{0} \\
\mathbf{0} & D_I \boldsymbol{\mathcal{L}}
\end{pmatrix},
\quad
\boldsymbol{\mathcal{J}} = 
\begin{pmatrix}
\mathbf{0} & \mathrm{diag}_\mu\left(\gamma - \tilde{\beta}_\mu S^*\right) \\[.2cm]
\mathbf{0} & \mathrm{diag}_\mu\left(\tilde{\beta}_\mu S^* - \gamma \right)
\end{pmatrix}.
\]
The Jacobian \( \boldsymbol{\mathcal{\tilde{J}}} \) is evaluated at the disease-free equilibrium, where each individual in every metanode is susceptible with probability \( S_{\mu} = S^* \), and no individual is infected, i.e., \( I_{\mu} = I^* = 0 \).
The block structure of \( \boldsymbol{\mathcal{\tilde{J}}} \) gives rise to an eigenvalue problem of the form 
\[
\boldsymbol{\mathcal{\tilde{J}}} \boldsymbol{\phi}^{(\eta)} = \lambda^{(\eta)} \boldsymbol{\phi}^{(\eta)},
\]
where \( \boldsymbol{\phi}^{(\eta)} = (\boldsymbol{\phi}_{S}^{(\eta)}, \boldsymbol{\phi}_{I}^{(\eta)})^\top \) denotes the eigenvector with components corresponding to the susceptible $S$ and infected $I$ species, respectively. Expanding the matrix product yields the following two-component system:
\begin{equation}
\begin{aligned}
\textit{First row:} & \; D_S \boldsymbol{\mathcal{L}} \boldsymbol{\phi}_S^{(\eta)} + \mathrm{diag}_\mu\left(\gamma - \tilde{\beta}_\mu S^*\right) \boldsymbol{\phi}_I^{(\eta)} = \lambda^{(\eta)} \boldsymbol{\phi}_S^{(\eta)}, \\[.1cm]
\textit{Second row:} & \; \left[D_I \boldsymbol{\mathcal{L}} - \mathrm{diag}_\mu\left(\gamma - \tilde{\beta}_\mu S^*\right) \right] \boldsymbol{\phi}_I^{(\eta)} = \lambda^{(\eta)} \boldsymbol{\phi}_I^{(\eta)}.
\end{aligned}
\label{eq:1_2_row}
\end{equation}

We now shift our focus to spectral properties of the Jacobian \( \boldsymbol{\mathcal{\tilde{J}}}_0 = \boldsymbol{\mathcal{\tilde{J}}} \big|_{\tilde{\beta}_\mu = \tilde{\beta}} \), obtained by setting all local transmission rates to a common value $\tilde{\beta}$; this corresponds to the zeroth-order approximation around the homogeneous case. These spectral characteristics form the foundation for the subsequent perturbative analysis. The results, derived in full in Appendix \ref{sec:LSA_App}, reveal two distinct kinds of eigenvalue–eigenvector pairs:

\renewcommand{\thetable}{I}
\setcounter{table}{0}
\begin{table}[h!]
\label{table}
\centering
\renewcommand{\arraystretch}{1.6}
\begin{tabular}{c|c|c}
\toprule
\emph{Kind} & \emph{Eigenvalue} & \emph{Eigenvector} \\
\midrule
\emph{I} & 
\begin{tabular}{c}
\( \lambda^{(\alpha)}_0 = D_S \Lambda^{(\alpha)} \) \\
\( \left(\alpha = 1, \dots, \Omega\right) \)
\end{tabular}
& 
\(
\boldsymbol{\phi}^{(\alpha)}_0 = 
\begin{pmatrix}
\boldsymbol{\Phi}^{(\alpha)} \\[.15cm]
\mathbf{0}
\end{pmatrix}
\) \\
\cmidrule{1-3}
\emph{II} & 
\begin{tabular}{c}
\( \lambda^{(\bar{\alpha})}_0 = \tilde{\beta} S^* - \gamma + D_I \Lambda^{(\alpha)} \) \\
\( \left(\bar{\alpha} = \Omega + 1, \dots, 2\Omega\right) \)
\end{tabular}
& 
\(
\boldsymbol{\phi}_0^{(\bar{\alpha})} = 
\begin{pmatrix}
\boldsymbol{\Phi}^{(\alpha)} \\[.15cm]
E_\alpha \boldsymbol{\Phi}^{(\alpha)}
\end{pmatrix}
\) \\
\bottomrule
\end{tabular}
\end{table}
\noindent where \( \Lambda^{(\alpha)} \) and \( \boldsymbol{\Phi}^{(\alpha)} \) denote the eigenvalues and eigenvectors of the Laplacian \( \boldsymbol{\mathcal{L}} \). The coefficient \( E_\alpha \) appearing in the second kind of eigenvectors is given by
\[
E_\alpha = \dfrac{(D_I - D_S) \Lambda^{(\alpha)} - (\gamma - \tilde{\beta} S^*)}{\gamma - \tilde{\beta} S^*}.
\]
In particular, when \( \Lambda^{(1)} = 0 \), we find \( E_1 = -1 \), leading to the explicit form \( \boldsymbol{\phi}_0^{(\Omega+1)} = (\boldsymbol{1}^\top, -\boldsymbol{1}^\top)^\top \).

Before proceeding with the spectral perturbative analysis, two conditions must be satisfied. First, to ensure the regularity of the expansion used in the linear stability analysis, spectral degeneracy must be avoided. This requires \( \tilde{\beta} S^* \neq \gamma \), as otherwise the eigenvalues \( \lambda^{(1)}_0 \) and \( \lambda^{(\Omega+1)}_0 \) coincide when \( \Lambda^{(1)} = 0 \), resulting in a degenerate eigenspace. In such a case, the Jacobian eigenvectors would not form a complete basis, rendering the second-order correction ill-defined. When this condition is met, the eigenvectors span the full space, and the expansion remains well-posed.

Second, the unperturbed system must be linearly stable with respect to the eigenvalues of the second kind. In particular, the leading eigenvalue associated with the zero Laplacian mode, \( \Lambda^{(1)} = 0 \), must satisfy \( \lambda^{(\Omega+1)}_0 = \tilde{\beta} S^* - \gamma < 0 \), ensuring decay of the corresponding mode. Since all other Laplacian eigenvalues are strictly negative, this condition guarantees that all second-kind modes decay. However, this requirement is necessary but not sufficient for global stability, as the mode \( \lambda^{(1)}_0 = 0 \) associated with the first kind (arising from \( \Lambda^{(1)} = 0 \)) remains marginal. As shown in Appendix \ref{sec:LSA_App}, once both conditions are satisfied, the perturbation projects orthogonally to the critical mode \( \boldsymbol{\phi}^{(1)} _0= \left(\boldsymbol{1}^\top, \boldsymbol{0}^\top\right)^\top \), ensuring that the resulting stability criterion is well posed.

Let us now initiate the perturbative analysis by considering the deviation introduced by structural heterogeneity, expressed symbolically as 
\(
\text{DBMF}_\mu = \text{DBMF} + \epsilon\, \delta \langle k\rangle_\mu,
\)
where \( \delta \langle k\rangle_\mu \) denotes the deviation of a community's mean degree from the global average \( \langle\langle k\rangle_\mu\rangle \). 
The perturbation will thus act on \( \boldsymbol{\mathcal{\tilde{J}}}_0 \) as the base operator:
\[
\boldsymbol{\mathcal{\tilde{J}}}_\epsilon = \boldsymbol{\mathcal{\tilde{J}}}_0 + \epsilon \boldsymbol{\mathcal{\tilde{J}}}_1.
\]
The perturbative matrix \( \boldsymbol{\mathcal{\tilde{J}}}_1 \) is given by:
\[
\boldsymbol{\mathcal{\tilde{J}}}_1 = \begin{pmatrix} 0 & \boldsymbol{\kappa} \\ 0 & -\boldsymbol{\kappa} \end{pmatrix},
\]
where \( \boldsymbol{\kappa} \) is a diagonal matrix with elements:
\[
\kappa_{\mu} = \frac{\beta \cdot \langle \langle k \rangle_{\mu} \rangle}{N} - \frac{\beta \cdot \langle k \rangle_{\mu}}{N}.
\] 
These diagonal elements satisfy the \emph{trace-free} condition:
\begin{align*}
    \sum_{\mu=1}^{\Omega} \kappa_{\mu} &= \sum_{\mu=1}^{\Omega} \left( \frac{\beta \cdot \langle \langle k \rangle_{\mu} \rangle}{N} - \frac{\beta \cdot \langle k \rangle_{\mu}}{N} \right)\\ 
    &= \frac{\beta}{N} \left( \sum_{\mu=1}^{\Omega} \langle \langle k \rangle_{\mu} \rangle - \sum_{\mu=1}^{\Omega} \langle k \rangle_{\mu} \right) = 0\,.
\end{align*}
This result holds regardless of the distribution of \( \langle k \rangle_{\mu} \), confirming that the perturbation matrix \( \boldsymbol{\kappa} \) remains trace-free under all configurations of \( \langle k \rangle_{\mu} \). 

Building on this framework, we will next explore higher-order perturbations in the Jacobian spectrum. 

\subsection{First Order Perturbation Analysis}

The first-order correction to the eigenvalues under the perturbation \( \boldsymbol{\mathcal{\tilde{J}}}_1 \) is given by:
\[
\lambda^{(\eta)}_1 = \boldsymbol{\phi}^{(\eta)\top}_0 \boldsymbol{\mathcal{\tilde{J}}}_1 \boldsymbol{\phi}^{(\eta)}_0, \quad \eta=1, \dots, 2\Omega\,,
\]
where the orthogonality of the Jacobian eigenvectors follows from Table \ref{table}, while their normalization is assumed for simplicity \footnote{It is important to note that the normalization of the Laplacian eigenvectors does not necessarily extend to the Jacobian eigenvectors. Instead, for the second set of eigenvalues, we must consider  
\(\boldsymbol{\phi}^{(\bar{\alpha})\top}_0 = \begin{pmatrix} \boldsymbol{\Phi}^{(\alpha)\top} & E_\alpha \boldsymbol{\Phi}^{(\alpha)\top} \end{pmatrix}/{\sqrt{1+E^2_\alpha}}\)}. We now examine the two types of eigenvalue-eigenvector pairs.\\

\emph{Case 1:} Substituting into the expression above:
\begin{align*}
       \lambda^{(\alpha)}_1 &= \begin{pmatrix} \boldsymbol{\Phi}^{(\alpha)\top} & \mathbf{0}^\top \end{pmatrix} \begin{pmatrix} \mathbf{0} & \boldsymbol{\kappa} \\[.1cm] \mathbf{0} & -\boldsymbol{\kappa} \end{pmatrix} \begin{pmatrix} \boldsymbol{\Phi}^{(\alpha)} \\[.1cm] \mathbf{0} \end{pmatrix}.\\
   &= \boldsymbol{\Phi}^{(\alpha)\top} \cdot \mathbf{0} = 0.
\end{align*}
Thus, in this case, the first-order correction vanishes.\\

\emph{Case 2:} In this case, we focus on the perturbation of the largest eigenvalue, corresponding to \( \Lambda^{(1)} = 0 \) of the Laplacian \(\boldsymbol{\mathcal{L}}\), as it is the most likely candidate to become positive and influence the system's stability. Substituting the normalized eigenvector $\boldsymbol{\phi}^{(\Omega + 1)}_0$, and noting that for this case \( E_1 = -1 \), the first-order correction is given by:
   \begin{align*}
   \lambda^{(\Omega + 1)}_1 &= \frac{1}{2}\begin{pmatrix} \boldsymbol{1^\top} & -\boldsymbol{1^\top} \end{pmatrix} \begin{pmatrix} \mathbf{0} & \boldsymbol{\kappa} \\ \mathbf{0} & -\boldsymbol{\kappa} \end{pmatrix} \begin{pmatrix} \boldsymbol{1} \\ -\boldsymbol{1} \end{pmatrix}\\
   &= - \boldsymbol{1^\top}\boldsymbol{\kappa} \boldsymbol{1}=0,
   \end{align*}
where the last equality follows from the trace-free property of \( \boldsymbol{\kappa} \). 
In conclusion, the first-order perturbation of the largest eigenvalue in the second set also vanishes. Although the first-order corrections for other eigenvalues in this set are generally nonzero, their zeroth-order values are much smaller than the leading one. Thus, to maintain consistency, we consider sufficiently small perturbations so that only the largest eigenvalue becomes unstable.

\subsection{Second Order Perturbation Analysis}

From the analysis above, the first-order correction \( \lambda^{(\eta)}_1 \) vanishes for the relevant modes, and as will be shown in the following, the second-order correction—particularly for the largest eigenvalue—is essential for assessing stability under perturbations.
Following the second-order perturbation formula, derived in Appendix \ref{sec:perturb}, we have  
\begin{equation}
\lambda_2^{(\eta)} = \sum_{\theta \neq \eta} \frac{\left( \boldsymbol{\phi}_0^{(\theta)\top} \boldsymbol{\mathcal{\tilde{J}}}_1 \boldsymbol{\phi}_0^{(\eta)} \right)^2}{\lambda_0^{(\eta)} - \lambda_0^{(\theta)}}, \quad \eta=1, \dots, 2\Omega\,,
\label{eq:IIord}
\end{equation}
assuming that the eigenvectors are normalized, i.e., \( \boldsymbol{\phi}_0^{(\eta)\top} \boldsymbol{\phi}_0^{(\theta)} = \delta_{\eta\theta} \). To analyze this expression, we consider two distinct cases.\\

\textit{Case 1:} In the first case, the numerator of the second-order correction term, \( \boldsymbol{\phi}_0^{(\theta)\top} \boldsymbol{\mathcal{\tilde{J}}}_1 \boldsymbol{\phi}_0^{(\alpha)} \), vanishes for all \( \theta \neq \alpha \). This follows from the first-order perturbation analysis, which established that \( \boldsymbol{\mathcal{\tilde{J}}}_1 \boldsymbol{\phi}_0^{(\alpha)} = \mathbf{0} \). As a result, \( \boldsymbol{\phi}_0^{(\theta)\top} \boldsymbol{\mathcal{\tilde{J}}}_1 \boldsymbol{\phi}_0^{(\alpha)} \) is always zero, leading to \( \lambda_2^{(\alpha)} = 0 \).\\

\textit{Case 2:} As previously discussed, to prevent the degenerate scenario dictated by the second-order perturbation formula \eqref{eq:IIord}, we adjust \( \gamma - \tilde{\beta} S^* \) to a small negative value close to zero. This adjustment ensures that the largest eigenvalue of \( \lambda^{(\bar{\alpha})}_0 \), namely \( \lambda_0^{(\Omega+1)} \), remains slightly smaller than the largest eigenvalue \( \lambda_0^{(1)} \) of \( \lambda^{(\alpha)}_0 \), while still exceeding the remaining eigenvalues in the spectrum. Consequently, the separation between \( \lambda^{(\bar{\alpha})}_0 \) and the remaining eigenvalues is maintained.

To analyze the second-order perturbation, we decompose the summation in \eqref{eq:IIord} into two parts. First, consider the term corresponding to \(\eta = 1\), where the denominator satisfies \( \lambda_0^{(\Omega+1)} - \lambda_0^{(1)} < 0 \). However, due to the structure of the eigenvectors, the numerator vanishes, i.e.,  
\(
\left( \boldsymbol{\phi}_0^{(1)\top} \boldsymbol{\mathcal{\tilde{J}}}_1 \boldsymbol{\phi}_0^{(\Omega+1)} \right)^2 = 0.
\)
This follows from:
\begin{align*}
\boldsymbol{\phi}_0^{(1)\top} \boldsymbol{\mathcal{\tilde{J}}}_1 \boldsymbol{\phi}_0^{(\Omega+1)} &= \frac{1}{\sqrt{2}} \begin{pmatrix} \boldsymbol{1}^\top \quad \boldsymbol{0}^\top \end{pmatrix} 
\begin{pmatrix} \mathbf{0} & \boldsymbol{\kappa} \\[.1cm] \mathbf{0} & -\boldsymbol{\kappa} \end{pmatrix} 
\begin{pmatrix} \boldsymbol{1} \\ -\boldsymbol{1} \end{pmatrix} \\ 
&= -\frac{1}{\sqrt{2}} \boldsymbol{1^\top} \boldsymbol{\kappa} \boldsymbol{1} = 0,
\end{align*}
which in turn follows from the trace-free property of \( \boldsymbol{\kappa} \). Consequently, this term does not contribute to \( \lambda_2^{(\bar{\alpha})} \).

Next, consider the case \( \eta \neq \{1, \Omega+1\} \), where the denominator \( \lambda_0^{(\Omega+1)} - \lambda_0^{(\eta)} > 0 \). Since the numerator \( \left( \boldsymbol{\phi}_0^{(\eta)\top} \boldsymbol{\mathcal{\tilde{J}}}_1 \boldsymbol{\phi}_0^{(\Omega+1)} \right)^2 \geq 0 \), this part of the summation may contribute positively to \( \lambda_2^{(\Omega+1)} \). Specifically, we have  
\[
\boldsymbol{\phi}_0^{(\eta)\top} \boldsymbol{\mathcal{\tilde{J}}}_1 \boldsymbol{\phi}_0^{(\Omega+1)} =
\begin{cases} 
-\dfrac{1}{\sqrt{2}}\boldsymbol{\Phi}^{(\alpha)\top} \boldsymbol{\kappa} \boldsymbol{1}, & \eta = 2, \dots, \Omega, \\[0.5cm]
\dfrac{E_\alpha - 1}{\sqrt{2\left(1+E^2_\alpha\right)}}\boldsymbol{\Phi}^{(\alpha)\top} \boldsymbol{\kappa} \boldsymbol{1}, & \eta\ \text{otherwise},
\end{cases}
\]
where in both cases above, \( \alpha = 2, \dots, \Omega \). All these terms are, in principle, nonzero since \( \boldsymbol{\Phi}^{(\alpha)\top} \neq \boldsymbol{1}^\top \), and their combination results in a strictly positive second-order correction, \( \lambda_2^{(\Omega+1)} > 0 \). 

In conclusion, we have shown that above-average-density metacommunities drive the system toward instability, leading to a global surge in infections.


\section{Localization and reduction approach to pattern prediction}
\label{sec:VI}

Although the previous analysis showed that densely connected communities can destabilize the system and trigger a global outbreak, it does not reveal the role of individual nodes in the spreading process or how infections are distributed across the metapopulation. To address this, we adopt a pattern formation approach, which predicts spatial patterns in second-order phase transitions—recently extended to networks~\cite{cross_pattern_2009, qian2025explosive, Siebert, symm_break, muolo_persistence_2024, padmore2020modelling, pranesh2024effect, luo2024relationship}. As seen in Fig.~\ref{fig:Fig1}, the bifurcation from the disease-free to the endemic state is continuous, making such analysis applicable. Near the critical point, the system undergoes small-amplitude deviations, remaining close to the linear regime given by \[\sum_{\eta=1}^{2\Omega} C_{\eta} e^{\lambda^{(\eta)} t} \boldsymbol{\phi}^{(\eta)}.
\] Thus, if a single unstable mode is present—specifically \( \lambda^{(\Omega+1)} \), as identified through spectral perturbation analysis—the corresponding eigenvector of the Jacobian, \( \boldsymbol{\phi}^{(\Omega+1)} \), serves as a predictive indicator for both the evolution and the eventual spatial distribution of the infection.

\begin{figure*}
    \includegraphics[width=\textwidth]{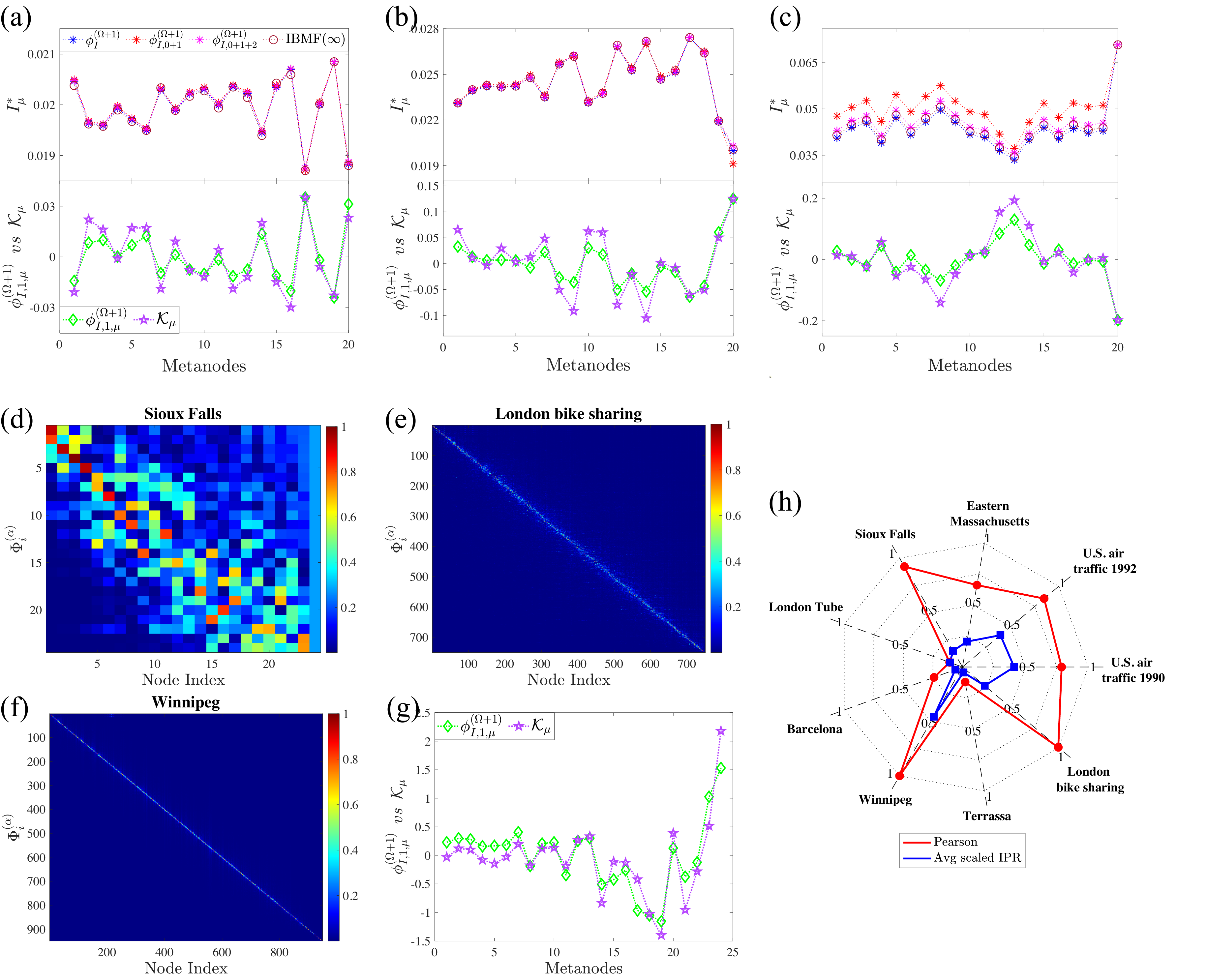}
\caption{The figure illustrates the distribution of metanodes along the horizontal axis, with the vertical axis divided into two sections. Panels (a)--(c) show synthetic benchmarks where the upper plots compare the asymptotic infection state $\mathbf{I}^{*}$ obtained from IBMF simulations with the critical eigenvector $\boldsymbol{\phi}^{(\Omega+1)}_{I}$ and its successive approximations, while the lower plots compare the deviation vector $\boldsymbol{\kappa}$ with the first-order correction $\boldsymbol{\phi}^{(\Omega+1)}_{I,1}$; the imposed condition $\tilde{\beta} S^{*} - \gamma = -0.01$ ensures distinct eigenvalues, and only $\lambda^{(\Omega+1)}$ is controlled to be positive. Panels (d)--(f) display eigenvector localization patterns for the Winnipeg traffic system, the London bike-sharing network, and the Eastern Massachusetts transportation network, highlighting the emergence of strongly localized modes in empirical structures. Panel (g) reports the direct comparison between the IBMF asymptotic infection state, the critical eigenvector, and its first-order correction for the Sioux Falls traffic network, using the same parameter setting as in panels (a)--(c) to establish consistency between synthetic and empirical results. Panel (h) summarizes localization properties across the empirical datasets through a radar plot that jointly displays the Pearson correlation and normalized IPR, with methodological details provided in Appendix \ref{sec:local_valid}. For the synthetic networks, we use the parameters $\Omega = 20$, $N = 50$, and $q = 0.5$, with probabilities drawn from a normal distribution of mean $0.5$ and standard deviations of $0.01$, $0.05$, and $0.08$. In both the synthetic and Sioux Falls cases, the dynamical parameters are $\beta = 0.5$, $D_S = 0.2$, and $D_I = 0.1$.}
    \label{fig:Fig3}
\end{figure*}

In Fig.~\ref{fig:Fig3}, the upper part of panels (a-c) compare the asymptotic infection levels \( I_\mu^* \) from IBMF simulations with the critical eigenvector \( \boldsymbol{\phi}^{(\Omega+1)} \) and its perturbative approximations. As the variance in community densities increases, differences between approximations grow more evident, yet the second-order correction closely tracks the asymptotic state, confirming the predictive power of the spectral framework. The lower part of the panels (a-c) show that the infection-related component of the first-order eigenvector correlates strongly with the mean degrees of the communities—highlighting that even a low-order spectral approximation captures how structural heterogeneity shapes epidemic outcomes. This same relationship is confirmed in panel (g) for the Sioux Falls road network, where the first-order correction likewise tracks community connectivity, demonstrating that the spectral approach extends robustly to empirical settings. Together, these results emphasize the role of Jacobian eigenvectors in linking connectivity patterns to the spatial distribution of infections.

Building on this intuition, we next focus on the localization of the Laplacian eigenvectors, which play a key role in shaping the spatial structure of infection patterns. The Laplacian matrix \( \boldsymbol{\mathcal{L}} \) admits a set of localized eigenvectors \( \boldsymbol{\phi}^{(\alpha)} \), whose components are significantly non-zero only on a very small number of nodes. These eigenvectors, when ordered appropriately, form the columns of the matrix \( \boldsymbol{\Phi} \), which itself appears nearly diagonal. This effect becomes stronger with increasing network randomness and, consequently, with the size of the network~\cite{mcgraw, nakao_loc, pastor2016distinct}. For further details on the localization properties, see Fig.~\ref{fig:Fig5} in \ref{sec:local_valid}, where a systematic analysis of decoupling and reduction approach is presented. In the following, we will derive the result specifically focused on the current problem.

We start by considering the eigenvalue problem \eqref{eq:1_2_row} expanding the eigenvalue and eigenvectors in terms of a small perturbation parameter \(\epsilon\), we write
\begin{align*}
    \lambda^{(\Omega+1)} &= \lambda^{(\Omega+1)}_0 + \epsilon \lambda^{(\Omega+1)}_1 + \mathcal{O}(\epsilon^2), \\
    \boldsymbol{\phi}^{(\Omega+1)}_S &= \boldsymbol{\phi}^{(\Omega+1)}_{S,0} + \epsilon \boldsymbol{\phi}^{(\Omega+1)}_{S,1} + \mathcal{O}(\epsilon^2), \\
    \boldsymbol{\phi}^{(\Omega+1)}_I &= \boldsymbol{\phi}^{(\Omega+1)}_{I,0} + \epsilon \boldsymbol{\phi}^{(\Omega+1)}_{I,1} +  \mathcal{O}(\epsilon^2).
\end{align*}
Additionally, we expand \(\tilde{\beta}_{\mu}\) as \(\tilde{\beta}_{\mu} = \tilde{\beta} + \epsilon \kappa_{\mu}\).
Substituting these expansions into the governing equations, at zeroth order \(\mathcal{O}(1)\), we obtain
\begin{align*}
    D_S \boldsymbol{\mathcal{L}} \boldsymbol{\phi}^{(\Omega+1)}_{S,0} + \left(\gamma - \tilde{\beta} S^*\right)\mathbb{I}_\Omega \boldsymbol{\phi}^{(\Omega+1)}_{I,0} &= \lambda^{(\Omega+1)}_0 \boldsymbol{\phi}^{(\Omega+1)}_{S,0}, \\
    \left[D_I \boldsymbol{\mathcal{L}} - \left(\gamma - \tilde{\beta} S^*\right) \mathbb{I}_\Omega\right] \boldsymbol{\phi}^{(\Omega+1)}_{I,0} &= \lambda^{(\Omega+1)}_0 \boldsymbol{\phi}^{(\Omega+1)}_{I,0}.
\end{align*}
Given that \( \lambda^{(\Omega+1)}_0 = -\gamma + \tilde{\beta} S^* \), \( \boldsymbol{\phi}^{(\Omega+1)}_{S,0} = \mathbf{1} \), and \( \boldsymbol{\phi}^{(\Omega+1)}_{I,0} = -\mathbf{1} \), and since  
\( \boldsymbol{\mathcal{L}} \mathbf{1} = 0 \), it follows that the zeroth-order contribution vanishes.
Proceeding to first order \(\mathcal{O}(\epsilon)\), we derive the system
\begin{align*}
    &D_S \boldsymbol{\mathcal{L}} \boldsymbol{\phi}^{(\Omega+1)}_{S,1} + \left(\gamma - \tilde{\beta} S^*\right) \mathbb{I}_\Omega \boldsymbol{\phi}^{(\Omega+1)}_{I,1} = S^*\boldsymbol{\kappa} \mathbf{1} + \lambda^{(\Omega+1)}_0 \boldsymbol{\phi}^{(\Omega+1)}_{S,1}, \\
    &\left[D_I \boldsymbol{\mathcal{L}} - \left(\gamma - \tilde{\beta} S^*\right) \mathbb{I}_\Omega\right] \boldsymbol{\phi}^{(\Omega+1)}_{I,1} = - S^* \boldsymbol{\kappa} \mathbf{1} + \lambda^{(\Omega+1)}_0 \boldsymbol{\phi}^{(\Omega+1)}_{I,1}.
\end{align*}
Summing these two equations eliminates the perturbation term \(S^*\boldsymbol{\kappa} \mathbf{1}\), and solving for \(\boldsymbol{\Phi}^{(\Omega+1)}_{I,1}\), yields
\begin{equation}
    D_I \boldsymbol{\mathcal{L}}\boldsymbol{\phi}^{(\Omega+1)}_{I,1} = - S^*\boldsymbol{\kappa} \mathbf{1}.
\end{equation}

From this equation $\boldsymbol{\phi}_{I,1}^{(\Omega+1)}$ and $-S^*\boldsymbol{\kappa} \mathbf{1}$ can be expressed as linear combinations of the eigenvectors of the Laplacian matrix. Thus, we have:
\[
D_I \boldsymbol{\mathcal{L}} \sum_{\alpha \neq 1} \tilde{C}_{\alpha}^1 \boldsymbol{\Phi}^{(\alpha)} = \sum_{\alpha} b_{\alpha} \boldsymbol{\Phi}^{(\alpha)},
\]
where the former follows the same spectral perturbation expansion of the eigenvectors as described in Appendix \ref{sec:perturb}, with the only difference that now \( \tilde{C}_{\alpha}^1 = C_{\alpha}^1 E_\alpha \). Proceeding it simplifies to
\[
\sum_{\alpha \neq 1} D_I \tilde{C}_{\alpha}^1 \Lambda^{(\alpha)}\boldsymbol{\Phi}^{(\alpha)} = \sum_{\alpha} b_{\alpha} \boldsymbol{\Phi}^{(\alpha)}\,,
\]
and using the orthogonality of eigenvectors, we deduce that
\[
b_\alpha =
  \begin{cases}
    0       & \alpha = 1,\\
    D_I \tilde{C}_{\alpha}^1 \Lambda^{(\alpha)}   & \alpha \neq 1\,,
  \end{cases}
\]
which follows from the fact that the vector $\boldsymbol{\kappa} \mathbf{1}$ has zero sum \footnote{That is, \( \mathbf{1}^\top \boldsymbol{\kappa} \mathbf{1} = \sum_\alpha b_\alpha\, \mathbf{1}^\top \boldsymbol{\Phi}^{(\alpha)}=0 \); this relation is automatically satisfied for all \( \alpha \neq 1 \) due to the orthogonality of the eigenvectors, so the only possible choice is \( b_1 = 0 \).}.
Consequently,
\[
\tilde{C}_{\alpha}^1=\frac{b_\alpha}{D_I\Lambda^{(\alpha)}}\,
\]
since \(\Lambda^{(\alpha )} < 0\) for \(\alpha \neq 1\).

Because not all eigenvectors are sufficiently localized, we assume a density distribution of the form \( \boldsymbol{\kappa} \mathbf{1} \). This approximation improves when aligned with the support of the dominant eigenvectors, specifically those indexed from \( \Pi + 1 \) to \( \Omega \). Applying these assumptions, we express
\[
-S^* \boldsymbol{\kappa} \mathbf{1} = \sum_{\alpha=1}^{\Pi} b_\alpha \boldsymbol{\Phi}^{(\alpha)} + \sum_{\alpha=\Pi+1}^{\Omega} b_\alpha \boldsymbol{\Phi}^{(\alpha)}.
\]
Then,
\[
b_\alpha =
  \begin{cases}
    0       & \alpha = 1,\dots,\Pi,\\
    -S^*\boldsymbol{\Phi}^{(\alpha)\top}  \boldsymbol{\kappa} \mathbf{1}  & \alpha=\Pi+1,\dots,\Omega\,,
  \end{cases}
\]
which we can rewrite in matrix form
\[
-S^* \boldsymbol{\kappa} \mathbf{1} = \tilde{\mathbf{\Phi}} \tilde{\mathbf{b}},
\]
where $\mathbf{\tilde{\Phi}}=[\boldsymbol{\Phi}^{(\Pi+1)},\dots,\boldsymbol{\Phi}^{(\Omega)}]$ and ${\mathbf{\tilde{b}}}=[b_{\Pi+1},\dots,b_{\Omega}]^\top$. At this point we will use the fact that $\mathbf{\tilde{\Phi}}$ is highly localized, with each eigenvector having a single nonzero entry. Thus, we obtain:
\begin{equation}
     S^* \kappa_\mu\approx
  \begin{cases}
    0       & \alpha = 1,\dots,\Pi,\\
    -b_{\alpha}{\Phi}^{(\alpha)}_{\mu}  & \alpha = \displaystyle\arg\max_{\alpha' \in \{\Pi+1, \dots, \Omega\}} \left|{\Phi}^{(\alpha')}_{\mu}\right|.
  \end{cases}
  \label{eq:29}
\end{equation}
That is, for each row index~$\mu$, $\alpha$ is chosen as the unique index in ${\Pi+1, \dots, \Omega}$ that maximizes $|{\Phi}^{(\alpha)}_{\mu}|$, and, due to the localization, each $\alpha$ is selected only once.
Since
\[
\sum_{\alpha \neq 1} D_I \tilde{C}_{\alpha}^1 \Lambda^{(\alpha)} \boldsymbol{\Phi}^{(\alpha)} = \sum_{\alpha=\Pi+1}^{\Omega} b_{\alpha} \boldsymbol{\Phi}^{(\alpha)},
\]
we derive
\begin{equation}
\tilde{C}_{\alpha}^1 =
  \begin{cases}
    0       & \alpha = 1,\dots,\Pi,\\
    \dfrac{b_{\alpha}}{D_I\Lambda^{(\alpha)}}   & \alpha=\Pi+1,\dots,\Omega.
  \end{cases}
    \label{eq:30}
\end{equation}
Thus,
\[
\boldsymbol{\phi}_{I,1}^{(\Omega+1)}=\sum_{\alpha>\Pi} \tilde{C}_{\alpha}^1 \boldsymbol{\Phi}^{(\alpha)}.
\]
Finally, combining together eqs. \eqref{eq:29} and \eqref{eq:30}, and making again use of the near diagonality of the eigenvectors matrix  we conclude that:
\begin{equation}
    \mathbf{\phi}_{I, 1, \mu}^{(\Omega+1)} \approx
  \begin{cases}
    0       & \alpha = 1,\dots,\Pi,\\
     -\dfrac{S^*\kappa_\mu }{D_I\Lambda^{(\alpha)}} & \alpha = \displaystyle\arg\max_{\alpha' \in \{\Pi+1, \dots, \Omega\}} \left|{\Phi}^{(\alpha')}_{\mu}\right|.
  \end{cases}
  \label{eq:local}
\end{equation}

Since, in pattern formation theory, the asymptotic infection profile satisfies \( I^*_\mu \sim \boldsymbol{\phi}_{I, \mu}^{(\Omega+1)} \)—the \( \mu \)-th entry of the infection component of the Jacobian’s critical eigenvector—the first-order correction in Eq.~\eqref{eq:local} establishes a direct link between the asymptotic infection state and the distribution of community densities. This analytical prediction is visually confirmed in Fig.~\ref{fig:Fig3}, where the lower panels compare the density deviation vector \( \boldsymbol{\kappa} \) with the first-order correction \( \boldsymbol{\phi}^{(\Omega+1)}_{I,1} \). The strong agreement highlights how structural heterogeneity governs the spatial distribution of infections.

To establish that our theory is not confined to synthetic benchmarks, we carry out a systematic validation on \emph{nine} empirical mobility and transportation networks spanning distinct geographies, scales, and data-generating processes. Across these real infrastructures we observe strongly localized eigenmodes consistent with the mechanisms predicted by our framework, indicating that the localization phenomenon is not an artifact of model construction. On the Sioux Falls traffic network, we perform a direct, out-of-sample comparison between the deviation vector $\boldsymbol{\kappa}$, and the first-order perturbative correction $\boldsymbol{\phi}^{(\Omega+1)}_{I,1}$, computed under the same parameterization used for the synthetic analyses; the close agreement among these three objects demonstrates that our analytic approximation accurately captures the empirical steady state. Finally, we quantify performance across all nine datasets using a joint radar summary of Pearson correlation and normalized inverse participation ratio (IPR), which corroborates both predictive accuracy and localization fidelity in heterogeneous real-world settings. Methodological details and robustness checks are provided in Appendix \ref{sec:local_valid}.



\section{Conclusions and discussion} \label{sec:conlus}

In this work, we addressed the question of how epidemic spreading unfolds in structured populations, focusing on the role of local contact networks within metapopulation dynamics — an aspect that, while previously studied through agent-based models, has rarely been explored using reduced or analytically tractable approaches. Some earlier studies have integrated local structure with large-scale mobility (e.g., \cite{watts2005multiscale, apolloni2014metapopulation, wang2022epidemic, granell2018epidemic}), but these typically assume homogeneous mixing within patches or rely on idealized mobility patterns. To bridge this gap, we developed a unified framework that explicitly accounts for both local heterogeneity and the hierarchical organization of the metapopulation, yielding a more realistic representation of spreading dynamics.

To make the complexity of local contact networks analytically tractable, we employed a degree-based mean-field approximation, which gives rise to a natural scaling of the contagion rate with the mean degree of each community. Despite its simplicity, this approximation accurately captures the behavior of the full system by preserving the effects of structural heterogeneity across metanodes. Using spectral perturbation theory, we show that heterogeneous contagion rates - unlike their homogeneous counterparts - amplify the influence of denser metanodes, which play a critical role in driving the global spread of infection. This highlights the disproportionate impact of communities with higher internal connectivity. Building on this, we proposed a method, to our knowledge, applied here for the first time, based on the localization of Laplacian eigenvectors, a known phenomenon in large random networks~\cite{mcgraw,nakao_loc}. This mapping between contagion rates and infection levels enables a predictive understanding of how structural heterogeneity shapes epidemic patterns. Empirically, the degree-based closure (DBMF$\mu$) closely reproduces the individual-based mean-field predictions on fixed real social contact networks, validating the approximation on real topologies. Moreover, across nine mobility/transportation systems, our perturbative–spectral predictors—the deviation vector $\boldsymbol{\kappa}$ and the first-order correction $\boldsymbol{\phi}^{(\Omega+1)}{I,1}$—accurately recover the observed steady states and localization signatures (high Pearson agreement and normalized IPR), confirming robust transfer from synthetic to real infrastructures.

Our work contributes to the broader network theory literature, intersecting with structured models such as multilayer~\cite{multilayer, multilayer_review}, modular~\cite{modular, communities}, networks of networks~\cite{nets_nets}, and interconnected networks~\cite{inter_nets}, where spreading dynamics have been extensively explored~\cite{multi_diffusion, multi_pattern, inter_epidemics}. Distinctively, it captures the coupling between local contact dynamics and global diffusion, resonating with the metaplex paradigm~\cite{metaplex}, which emphasizes integration of local structure with population-level processes. We formulate a hierarchical epidemic model that incorporates intra-community heterogeneity within a metapopulation setting, while offering analytical insights relevant to a broad class of structured population models—including age~\cite{cushing1998introduction, auger2008structured} and spatially structured~\cite{ostfeld2005spatial, doekes2024multiscale} approaches. Altogether, our findings highlight the need to reconcile local complexity with large-scale connectivity, offering a tractable and generalizable framework for studying multiscale epidemic dynamics.


\begin{acknowledgments}
M.A. acknowledges support from the SEED grant from the FSU Council on Research and Creativity SEED grant \textit{Structure and dynamics of nonnormal networks}.
\end{acknowledgments}


\appendix

\section{Degree-Based Mean-Field Approximation}
\label{sec:DBMF_App}

In this section, we will reduce the original IBMF model to a reaction-diffusion system using a degree-based mean-field approach which is a powerful approximation technique widely used to study networked complex systems~\cite{newman_book}. By averaging the high-dimensional interactions between individual nodes, it effectively simplifies the analysis of processes such as epidemic spreading and opinion dynamics \cite{pastor2001epidemic_1, pastor2001epidemic_2, qian2025explosive}. This method assumes that nodes with similar properties (e.g. degree or state) exhibit statistically homogeneous behavior, enabling the modeling of collective dynamics while mitigating computational complexity. 

To simplify the governing equations of the IBMF model, we define the aggregated quantities \( S_{\mu} = \sum_{i} S_{i,\mu}\) and \(I_{\mu} = \sum_{i} I_{i,\mu}, \) that represent the probability that individuals present in metanode \( \mu \) are susceptible or infected, respectively.
By substituting these definitions into the node-level equations in Eq.~(1) of the main text, we obtain the following form of the aggregated dynamics:
\begin{equation}
\begin{aligned}
    \dot{S}_{\mu} &= -\beta \sum_{i,j} A^{(\mu)}_{ij} S_{i,\mu} I_{j,\mu} + \gamma I_{\mu} + D_S \sum_{\nu} \mathcal{L}_{\mu \nu} S_{\nu}, \\
    \dot{I}_{\mu} &= \beta \sum_{i,j} A^{(\mu)}_{ij} S_{i,\mu} I_{j,\mu} - \gamma I_{\mu} + D_I \sum_{\nu} \mathcal{L}_{\mu \nu} I_{\nu}.
\end{aligned}
\end{equation}
To further analyze the contagion term, \(\sum_{i,j} A^{(\mu)}_{ij} S_{i,\mu} I_{j,\mu}\), we assume that the individual node-level variables \(S_{i,\mu}\) and \(I_{j,\mu}\) can be decomposed into their mean-field components and fluctuations:
\[
S_{i,\mu} = \langle S \rangle_{\mu} + \delta S_i, \quad I_{j,\mu} = \langle I \rangle_{\mu} + \delta I_j,
\]
where \(\langle S \rangle_{\mu} = S_{\mu} / N\) and \(\langle I \rangle_{\mu} = I_{\mu} / N\) represent the mean states for metanode and \(\delta S_i\), \(\delta I_j\) are the respective deviations from the mean. Substituting these expressions into the interaction term, and after expanding and simplifying, we obtain the following:
\begin{align*}
\beta \sum_{i,j} & A^{(\mu)}_{ij} S_{i,\mu} I_{j,\mu} = \beta \langle S \rangle_{\mu} \left(\langle I \rangle_{\mu} \sum_{i,j} A^{(\mu)}_{ij} +  \sum_{i,j} A^{(\mu)}_{ij} \delta I_j\right)\\ &+ \beta \left( \langle I \rangle_{\mu} \sum_{i,j} A^{(\mu)}_{ij} \delta S_i + \sum_{i,j} A^{(\mu)}_{ij} \delta S_i \delta I_j\right).    
\end{align*}

The degree-based approximation we use here is based on the empirical assumption that the amount density associated with a given node scales with its degree, $x_i \sim k_i$. In addition, throughout this paper we will assume a symmetric and narrow distribution $P(k)$ of the contact network $\mathbf{A}$~\footnote{In our recent work \cite{qian2025explosive}, we have shown that the validity of this approximation extends beyond the standard assumptions, including cases with long-tailed degree distributions such as scale-free networks.}. Based on these considerations, the last term in the expression above, which involves \( \delta S_i \delta I_j \), represents a higher-order nonlinear correction, and when deviations are small, it can be neglected under the mean-field approximation.
Furthermore, for the adjacency matrix, we note that
\(
\sum_{i,j} A^{(\mu)}_{ij} = \sum_{i} k_i^{\mu} = \sum_{j} k_j^{\mu}= N \langle k \rangle_{\mu}
\)
where \(\langle k \rangle_{\mu}\) is the average degree of nodes in metanode \(\mu\). Collectively, these lead to

\begin{widetext} 
\begin{align*}
\beta \sum_{i,j} A^{(\mu)}_{ij} S_{i,\mu} I_{j,\mu} &\approx \beta N \langle k \rangle_{\mu} \langle S \rangle _{\mu} \langle I \rangle_{\mu}
+ \beta \langle S \rangle_{\mu} \sum_{j} k_j^{\mu} \delta I_j + \beta \langle I \rangle_{\mu} \sum_{i} k_i^{\mu} \delta S_i \\ 
&= \beta N \langle k \rangle_{\mu} \langle S \rangle _{\mu} \langle I \rangle_{\mu} 
+ \beta \langle S \rangle _{\mu} \langle k \rangle _{\mu}\sum_i\left(\delta S_i + \delta I_i\right) 
+ \beta \langle S \rangle _{\mu} \sum_i \delta k_i^{\mu}\left(\delta S_i + \delta I_i\right)\,.\\
\end{align*}
\end{widetext} 

The terms containing \(\delta k_i^{\mu} \delta I_i\) and \(\delta k_i^{\mu} \delta S_i\) are omitted as they are higher-order terms. Furthermore, since \(\delta k_i^{\mu}\), \(\delta S_i\), and \(\delta I_i\) exhibit symmetry, any terms involving \(\sum_i \delta S_i\), \(\sum_i \delta I_i\), and \(\sum_i \delta k_i^{\mu}\) will vanish.
In conclusion, we have

\[
\beta \sum_{i,j} A^{(\mu)}_{ij} S_{i,\mu} I_{j,\mu} \approx \beta N \langle k \rangle_{\mu} \langle S \rangle_{\mu} \langle I \rangle_{\mu}.
\]

Rewriting in terms of the aggregated quantities, we define an effective transmission rate \(\tilde{\beta}_{\mu} = \beta \langle k \rangle_{\mu}/N\), and obtain the system:

\begin{equation}
\label{eq:DBMF_mu}
\begin{aligned}
\dot{S}_{\mu} &= -\tilde{\beta}_{\mu} S_{\mu} I_{\mu} + \gamma I_{\mu} + D_S \sum_{\nu} \mathcal{L}_{\mu \nu} S_{\nu}, \\
\dot{I}_{\mu} &= \tilde{\beta}_{\mu} S_{\mu} I_{\mu} - \gamma I_{\mu} + D_I \sum_{\nu} \mathcal{L}_{\mu \nu} I_{\nu}, \quad &\forall \mu\,.
\end{aligned}
\end{equation}
We refer to system Eq.~\eqref{eq:DBMF_mu} as the Degree-Based Mean-Field (\( \text{DBMF}_\mu \)) approximation within the broader metapopulation framework, which incorporates both diffusion across layers and locally averaged interactions, yielding a tractable description of the macroscopic dynamics while retaining key features of the underlying contact processes.


\section{Spectral perturbation theory}
\label{sec:perturb}

We present the general framework for spectral perturbation theory, focusing primarily on second-order corrections \cite{franklin2012matrix, horn2012matrix}. Starting from a symmetric matrix \( \boldsymbol{\mathcal{M}}_0 \) with known eigenvalues and eigenvectors, we introduce a small perturbation \( \epsilon \boldsymbol{\mathcal{M}}_1 \), such that the perturbed matrix can be written as:
\[
\boldsymbol{\mathcal{M}}_\epsilon = \boldsymbol{\mathcal{M}}_0 + \epsilon \boldsymbol{\mathcal{M}}_1
\]
For symmetric matrices \( \boldsymbol{\mathcal{M}}_0 \), the eigenvectors are orthogonal and can be denoted as \( \boldsymbol{\phi}_0^{(\eta)} \). Our objective is to compute the changes in eigenvalues and eigenvectors of \( \boldsymbol{\mathcal{M}}_\epsilon \), with emphasis on second-order corrections. The \emph{continuity theorem of eigenvalues} guarantees that the eigenvalues of \( \boldsymbol{\mathcal{M}}_\epsilon \) vary continuously with \( \epsilon \), forming the basis for the perturbative expansion \cite{horn2012matrix}.

The eigenvalues \( \lambda^{(\eta)} \) and eigenvectors \( \boldsymbol{\phi}^{(\eta)} \) of \( \boldsymbol{\mathcal{M}}_\epsilon \) are expanded as power series in \( \epsilon \):

\[
\lambda^{(\eta)} = \lambda_0^{(\eta)} + \epsilon \lambda_1^{(\eta)} + \epsilon^2 \lambda_2^{(\eta)} + \cdots
\]
\[
\boldsymbol{\phi}^{(\eta)} = \boldsymbol{\phi}_0^{(\eta)} + \epsilon \boldsymbol{\phi}_1^{(\eta)} + \epsilon^2 \boldsymbol{\phi}_2^{(\eta)} + \cdots
\]

\subsection{First-Order Corrections}

The first-order correction to the eigenvalue is given by:
\begin{equation}
    \lambda_1^{(\eta)} = \frac{\boldsymbol{\phi}_0^{(\eta)^\top} \boldsymbol{\mathcal{M}}_1 \boldsymbol{\phi}_0^{(\eta)}}{\boldsymbol{\phi}_0^{(\eta)^\top} \boldsymbol{\phi}_0^{(\eta)}}.
    \label{eq:P_eigval_1}
\end{equation}
The corresponding correction to the eigenvector is:
\begin{equation}
\boldsymbol{\phi}_1^{(\eta)} = \sum_{\theta \neq \eta} C_\theta^1 \boldsymbol{\phi}_0^{(\theta)}
    \label{eq:P_eigvec_1}
\end{equation}
where the coefficients \( C_\theta^1 \) are:
\[
C_\theta^1 = \frac{\mathbf{\phi}_0^{(\theta)\top} \boldsymbol{\mathcal{M}}_1 \boldsymbol{\phi}_0^{(\eta)}}{\left(\lambda_0^{(\eta)} - \lambda_0^{(\theta)}\right) \boldsymbol{\phi}_0^{(\theta)\top} \boldsymbol{\phi}_0^{(\theta)}}.
\]

\subsection{Second-Order Corrections}

The second-order correction to the eigenvalue \( \lambda_2^{(\theta)} \) is:

\begin{equation}
    \lambda_2^{(\eta)} = \frac{\boldsymbol{\phi}_0^{(\eta)\top} \boldsymbol{\mathcal{M}}_1 \boldsymbol{\phi}_1^{(\eta)}}{\boldsymbol{\phi}_0^{(\eta)\top} \boldsymbol{\phi}_0^{(\eta)}}.
    \label{eq:P_eigval_2}
\end{equation}
Substituting \( \boldsymbol{\phi}_1^{(\eta)} \) into the expression yields:
\[
\lambda_2^{(\eta)} = \sum_{\theta \neq \eta} \frac{C_\theta^1 \boldsymbol{\phi}_0^{(\eta)\top} \boldsymbol{\mathcal{M}}_1 \boldsymbol{\phi}_0^{(\theta)}}{\boldsymbol{\phi}_0^{(\eta)\top} \boldsymbol{\phi}_0^{(\eta)}}.
\]
Utilizing the symmetry property of the matrix:
\[
\boldsymbol{\phi}_0^{(\eta)\top} \boldsymbol{\mathcal{M}}_1 \boldsymbol{\phi}_0^{(\theta)} = \boldsymbol{\phi}_0^{(\theta)\top} \boldsymbol{\mathcal{M}}_1 \boldsymbol{\phi}_0^{(\eta)},
\]
the second-order correction becomes:
\[
\lambda_2^{(\eta)} = \sum_{\theta \neq \eta} \frac{\left( \boldsymbol{\phi}_0^{(\theta)\top} \boldsymbol{\mathcal{M}}_1 \boldsymbol{\phi}_0^{(\eta)} \right)^2}{\left(\lambda_0^{(\eta)} - \lambda_0^{(\theta)}\right) \boldsymbol{\phi}_0^{(\theta)\top} \boldsymbol{\phi}_0^{(\theta)}\boldsymbol{\phi}_0^{(\eta)\top} \boldsymbol{\phi}_0^{(\eta)}}.
\]
Assuming normalized eigenvectors, the expression above simplifies further. The sign of the correction depends on the difference \( \lambda_0^{(\eta)} - \lambda_0^{(\theta)} \).

The coefficient for the second-order eigenvector correction is given by:
\[
C_\theta^2=\dfrac{{\boldsymbol{\phi}_{0}^{(\theta)\top}}\boldsymbol{\mathcal{M}}_1\boldsymbol{\phi}_{1}^{(\eta)}-\lambda_1^{(\eta)}\sum_{\xi\neq\eta}C_\xi^{1}{\boldsymbol{\phi}_{0}^{(\theta)\top}}\boldsymbol{\phi}_{0}^{(\xi)}}{\left(\lambda_0^{(\theta)}-\lambda_0^{(\eta)}\right){\boldsymbol{\phi}_{0}^{(\theta)\top}}\boldsymbol{\phi}_{0}^{(\theta)}}\,.
\]
Since the first-order eigenvector correction \( \boldsymbol{\phi}_1^{(\eta)} \) is given by Eq.~\eqref{eq:P_eigvec_1}, we substitute it into the expression for \( C_\theta^2 \):
\[
C_\theta^2 = \frac{\boldsymbol{\phi}_{0}^{(\theta)\top} \left(\boldsymbol{\mathcal{M}}_1 - \lambda_1^{(\eta)} \mathbb{I}_\Omega\right) \boldsymbol{\phi}_{1}^{(\eta)}}
{\left(\lambda_0^{(\theta)}-\lambda_0^{(\eta)}\right){\boldsymbol{\phi}_{0}^{(\theta)\top}}\boldsymbol{\phi}_{0}^{(\theta)}}\,.
\]
The second-order correction to the eigenvector then reads:
\begin{equation}
    \boldsymbol{\phi}_2^{(\eta)} = \sum_{\theta \neq \eta} C_\theta^2 \boldsymbol{\phi}_0^{(\theta)}\,.
    \label{eq:P_eigvec_2}
\end{equation}


\section{Linear Stability Analysis of the DBMF model}
\label{sec:LSA_App}

\subsection{\texorpdfstring{Spectral properties of the Jacobian matrix $\boldsymbol{\mathcal{\tilde{J}}}_0$}{Spectral properties of the Jacobian matrix J0}}

To analyze the stability of the DBMF system, we begin by assuming a steady-state solution and introduce small perturbations around the equilibrium. Specifically, we express the system variables as
\[
S_{\mu} = S^{*}_{\mu} + \delta S_{\mu}, \quad I_{\mu} = I^{*}_{\mu} + \delta I_{\mu}, \quad \forall \mu
\]
where the steady-state values satisfy \( S^{*}_{\mu} = S^{*} \) and \( I^{*}_{\mu} = 0 \). Substituting these expressions into Eq.~\eqref{eq:DBMF_mu} and performing a linear stability analysis, we obtain:
\begin{equation*}
\label{eq:2}
\begin{aligned}
    \dot{\delta S}_{\mu} &= \left(\gamma - \tilde{\beta} S^{*}\right) \delta I_{\mu} + D_S \sum_{\nu} \mathcal{L}_{\mu \nu} \delta S_{\nu},\\ 
    \dot{\delta I}_{\mu} &= \left(\tilde{\beta} S^{*} - \gamma\right) \delta I_{\mu} + D_I \sum_{\nu} \mathcal{L}_{\mu \nu} \delta I_{\nu}, \quad &\forall \mu\,.
\end{aligned}
\end{equation*}
We further use the eigenvalue decomposition of the Laplacian
\(
\sum_{\nu=1}^\Omega \mathcal{L}_{\mu \nu} \Phi^{(\alpha)}_\nu = \Lambda^{(\alpha)} \Phi^{(\alpha)}_\mu,
\)
where \(\Phi^{(\alpha)}_\mu\) is the Laplacian eigenvector entry corresponding to the eigenvalue \(\Lambda^{(\alpha)}\). The eigenvalues of the Laplacian matrix \( \boldsymbol{\mathcal{L}} \) satisfy the decreasing ordering 
\(
\Lambda^{(\Omega)} \leq \dots \leq \Lambda^{(2)} \leq \Lambda^{(1)} = 0,
\)
where \( \Lambda^{(1)} \) is the corresponding eigenvalue of the uniform eigenvector \( \boldsymbol{\Phi}^{(1)}=\mathbf{1}\), and all others are non-positive. The perturbations \(\delta S_\mu\) and \(\delta I_\mu\) can then be expressed as follows:
\begin{equation}
\delta S_\mu = \sum_{\alpha=1}^\Omega b_{\alpha} e^{\lambda_0^{(\alpha)} t} \Phi_{\mu}^{(\alpha)}, \quad
\delta I_\mu = \sum_{\alpha=1}^\Omega c_{\alpha} e^{\lambda_0^{(\alpha)} t} \Phi_{\mu}^{(\alpha)},
\end{equation}
where \( \lambda^{(\alpha)}_0 \) is the growth rate, while \( b_\alpha \) and \( c_\alpha \) are constants set by the initial conditions.
Substituting these expressions into the linearized equations above decouples the system, yielding for each index \( \alpha \) the following condition for the existence of a solution, where \( \lambda^{(\alpha)}_0 \) now represents the Jacobian eigenvalue:
\begin{eqnarray*}
\det \begin{pmatrix}
D_S \Lambda^{(\alpha)} - \lambda^{(\alpha)}_0 & \gamma - \tilde{\beta} S^{*} \\[.15cm]
0 & \tilde{\beta} S^{*} - \gamma + D_I \Lambda^{(\alpha)} - \lambda^{(\alpha)}_0
\end{pmatrix} = 0.\nonumber\\
\end{eqnarray*}
Based on this equation, we distinguish between two kinds of eigenvalues:
\begin{equation}
\lambda^{(\alpha)}_0 = D_S \Lambda^{(\alpha)}, \quad
\lambda^{(\bar{\alpha})}_0 = \tilde{\beta} S^{*} - \gamma + D_I \Lambda^{(\alpha)},
\label{eq:eigvals}
\end{equation}
where \( \alpha \) ranges from \( 1 \) to \( \Omega \), while \( \bar{\alpha} \) spans \( \Omega + 1 \) to \( 2\Omega \), identifying the second type of eigenvalues. For compactness, we introduce \( \eta \in \{1, \dots, 2\Omega\} \), so that \( \lambda^{(\eta)}_0 \) represents all eigenvalues of the zeroth-order Jacobian matrix.

\begin{figure}[t]
\centering
\includegraphics[width=0.5\textwidth]{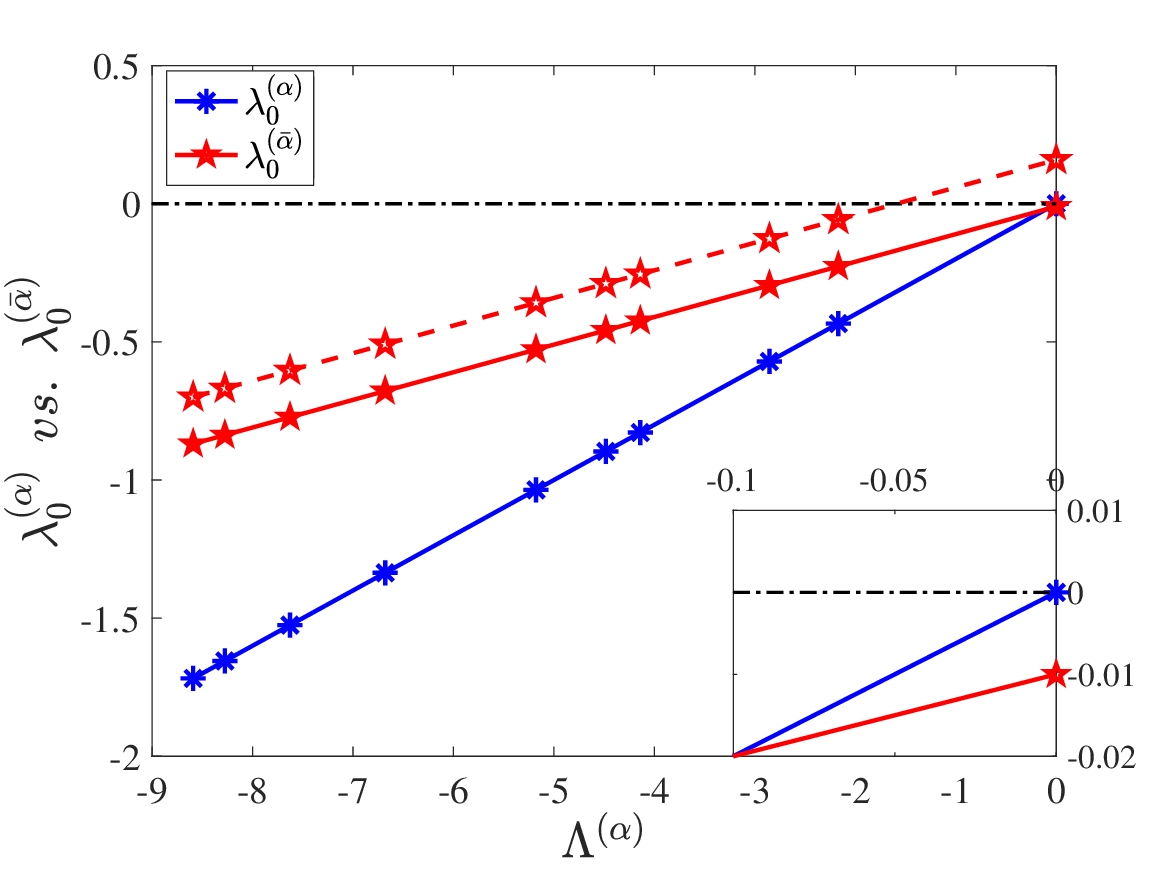}
\caption{The graph illustrates the two types of eigenvalues, \( \lambda^{(\alpha)}_0 \) and \( \lambda^{(\bar{\alpha})}_0 \), both corresponding to \( \Lambda^{(\alpha)} \) but differing in slope, associated with \( D_S \) and \( D_I \), respectively. Notably, only \( \lambda^{(\bar{\alpha})}_0 \) can become unstable or strictly stable, as illustrated by the dashed line. The inset provides a zoomed-in view, confirming that in our setting, the largest eigenvalue for \( \lambda^{(\bar{\alpha})}_0 \) is chosen as \( \tilde{\beta} S^{*} - \gamma = -0.01 \) to prevent spectral degeneracy.}
\label{fig:Fig2}
\end{figure}

These considerations are schematically illustrated in Fig.~\ref{fig:Fig2}, which shows both classes of eigenvalues. For the system to be linearly unstable, we require \( \lambda^{(\bar{\alpha})}_0 > 0 \), which leads to the condition:
\begin{equation}
\tilde{\beta} S^{*} - \gamma > 0.
\label{eq:insta}
\end{equation}
Among them, only \( \lambda^{(\bar{\alpha})}_0 \) can destabilize the system, as shown by Eq.~\eqref{eq:eigvals}, and indicated by the dashed red line. This inequality implies that the effective transmission rate \( \tilde{\beta} \), the susceptible steady-state fraction \( S^{*} \), or alternatively the recovery rate \( \gamma \), must cross a critical threshold to induce instability. Specifically, instability sets in when \( \gamma < \gamma_c := \tilde{\beta} S^{*} \), marking the onset of dynamic transitions in the system. This transition corresponds to the prediction by the DBMF of the epidemic threshold, as shown by the blue curve in Fig.~2 of the main text.

To investigate how the dynamics change when structural heterogeneity is introduced—i.e., under DBMF\(_\mu\)—we will perform a perturbative expansion around the DBMF solution. For this approach to be valid, the unperturbed system must be linearly stable; in particular, \( \lambda^{(\bar{\alpha})}_0 < 0 \) is a necessary condition (as we will show in the following, though not sufficient on its own).
To perform the expansion, spectral degeneracy must be avoided. This requires \( \tilde{\beta} S^{*} - \gamma \neq 0 \), as setting it to zero would imply \( \lambda^{(1)}_0 = \lambda^{(\Omega + 1)}_0 = 0 \) when \( \Lambda^{(1)} = 0 \), leading to degenerate eigenvalues. The inset of Fig.~\ref{fig:Fig2} shows the resulting eigenvalue separation, which ensures regularity. As shown in Appendix \ref{sec:perturb}, such degeneracy renders the second-order perturbation ill-defined; a detailed discussion follows in the next section.

A deeper understanding of the role of \( \text{DBMF}_\mu \) in the system's dynamics requires analyzing the spectral properties of the Jacobian matrix, not only through its eigenvalues but also its eigenvectors, which govern the system’s modal decomposition and reveal how perturbations evolve and grow across different dynamical regimes.
To this end, we now shift our focus to the full system rather than its decoupled components, analyzing the structure of the Jacobian \( \boldsymbol{\mathcal{\tilde{J}}}_0 \), a \( 2\Omega \times 2\Omega \) block matrix defined as
\[
\boldsymbol{\mathcal{\tilde{J}}}_0 = \boldsymbol{\mathcal{{D}}} + \boldsymbol{\mathcal{J}},
\]
where  
\[
\boldsymbol{\mathcal{{D}}} = 
\begin{pmatrix}
D_S \boldsymbol{\mathcal{L}} & \mathbf{0} \\
\mathbf{0} & D_I \boldsymbol{\mathcal{L}}
\end{pmatrix},
\quad
\boldsymbol{\mathcal{J}} = 
\begin{pmatrix}
\mathbf{0} & \left(\gamma - \tilde{\beta} S^*\right)\mathbb{I}_\Omega \\[.2cm]
\mathbf{0} & \left(\tilde{\beta} S^* - \gamma\right) \mathbb{I}_\Omega
\end{pmatrix}.
\]
The block structure of \( \boldsymbol{\mathcal{\tilde{J}}} \) gives rise to two sets of eigenvalues, with corresponding eigenvectors of the form  
\[
\boldsymbol{\phi}^{(\eta)}_0 = \begin{pmatrix} \boldsymbol{\phi}_{S,0}^{(\eta)} \\[.15cm] \boldsymbol{\phi}_{I,0}^{(\eta)} \end{pmatrix},
\]
where \( \boldsymbol{\phi}_{S,0}^{(\eta)} \) and \( \boldsymbol{\phi}_{I,0}^{(\eta)} \) correspond to the subspaces of \( \delta S \) and \( \delta I \), respectively. These components satisfy different possible configurations depending on the spectral properties of \( \boldsymbol{\mathcal{L}} \) and the interplay between \( \boldsymbol{\mathcal{D}} \) and \( \boldsymbol{\mathcal{{J}}} \). Since we already assumed the eigenvalues to be  distinct, the corresponding eigenvectors must be linearly independent, ensuring they form a complete basis. Consequently, either \( \boldsymbol{\phi}_{S,0}^{(\eta)} \) or \( \boldsymbol{\phi}_{I,0}^{(\eta)} \) may vanish, but both cannot be zero simultaneously, as this would violate linear independence.

To determine the corresponding eigenvectors, we consider the eigenvalue equation \( \boldsymbol{\mathcal{\tilde{J}}}_0 \boldsymbol{\phi}^{(\eta)}_0 = \lambda^{(\eta)}_0 \boldsymbol{\phi}^{(\eta)}_0 \), leading to  
\[
\begin{pmatrix}
D_S \boldsymbol{\mathcal{L}}  & \left(\gamma - \tilde{\beta} S^*\right) \mathbb{I}_\Omega \\[.15cm]
\mathbf{0} & D_I \boldsymbol{\mathcal{L}} - \left(\gamma - \tilde{\beta} S^*\right) \mathbb{I}_\Omega
\end{pmatrix}
\begin{pmatrix}
\boldsymbol{\phi}_{S,0}^{(\eta)} \\[.15cm]
\boldsymbol{\phi}_{I,0}^{(\eta)}
\end{pmatrix}
=
\lambda^{(\eta)}_0
\begin{pmatrix}
\boldsymbol{\phi}_{S,0}^{(\eta)} \\[.15cm]
\boldsymbol{\phi}_{I,0}^{(\eta)}
\end{pmatrix}.
\]
Expanding the matrix product, we obtain the following equations for the two components:
\begin{equation}
\begin{aligned}
\textit{First row:} & \,\,\,\, D_S \boldsymbol{\mathcal{L}} \boldsymbol{\phi}_{S,0}^{(\eta)} + \left(\gamma - \tilde{\beta} S^*\right)\mathbb{I}_\Omega \boldsymbol{\phi}_{I,0}^{(\eta)} = \lambda^{(\eta)}_0 \boldsymbol{\phi}_{S,0}^{(\eta)}, \\[.1cm]
\textit{Second row:} & \,\,\,\, \left[D_I \boldsymbol{\mathcal{L}} - \left(\gamma - \tilde{\beta} S^*\right) \mathbb{I}_\Omega\right] \boldsymbol{\phi}_{I,0}^{(\eta)} = \lambda^{(\eta)}_0 \boldsymbol{\phi}_{I,0}^{(\eta)}.
\end{aligned}
\label{eq:1_2_row}
\end{equation}

\subsubsection{Case 1: Eigenvalues \( \lambda^{(\alpha)}_0 = D_S \Lambda^{(\alpha)} \)}

Substituting \(\lambda^{(\alpha)}_0 = D_S \Lambda^{(\alpha)}\) into the first row and after rearranging terms
 gives:
\[\
\left(\tilde{\beta} S^* - \gamma\right)\, \boldsymbol{\phi}_{I,0}^{(\alpha)} = D_S \left(\boldsymbol{\mathcal{L}}\,\boldsymbol{\phi}_{S,0}^{(\alpha)} - \Lambda^{(\alpha)}\boldsymbol{\phi}_{S,0}^{(\alpha)}\right) .
\]
The structure of the right-hand side naturally suggests selecting \( \boldsymbol{\phi}_{S,0}^{(\alpha)} \) as an eigenvector of the Laplacian, that is, \( \boldsymbol{\phi}_{S,0}^{(\alpha)} = \boldsymbol{\Phi}^{(\alpha)} \). In this case, the equation reduces to:
\[
\left(\tilde{\beta} \, S^* - \gamma\right) \boldsymbol{\phi}_{I,0}^{(\alpha)} = \mathbf{0}.
\]
Given that \( \tilde{\beta} \, S^* - \gamma \neq 0 \), we conclude that:
\[
\boldsymbol{\phi}_{I,0}^{(\alpha)} = \mathbf{0}.
\]
Recall that \( \boldsymbol{\phi}_{S,0}^{(\alpha)} = \mathbf{0} \) is excluded, since both components of the eigenvector would vanish, contradicting the assumption of nontriviality.

The choices for \( \boldsymbol{\phi}_{S,0}^{(\alpha)} \) and \( \boldsymbol{\phi}_{I,0}^{(\alpha)} \) are consistent, as they ensure the eigenvalue equation remains valid. In fact, substituting \( \boldsymbol{\phi}_{I,0}^{(\alpha)} = \mathbf{0} \) into the second row confirms this consistency and gives:
\[
\left[D_I \boldsymbol{\mathcal{L}}  - \left(\gamma - \tilde{\beta} S^*\right) \mathbb{I}_\Omega\right] \mathbf{0} = \lambda_0^{(\alpha)} \mathbf{0}.
\]
This equation is trivially satisfied, imposing no further conditions.
Thus, summarizing, for the eigenvalues \( \lambda_0^{(\alpha)} = D_S \Lambda^{(\alpha)} \), the corresponding eigenvectors of \( \tilde{\mathbf{J}}_0 \) are given by:
\begin{equation}
    \boldsymbol{\phi}^{(\alpha)}_0 = \begin{pmatrix} \boldsymbol{\Phi}^{(\alpha)} \\[.1cm] \mathbf{0} \end{pmatrix}\,,
\label{eq:eigvec_1}
\end{equation}
which are unique (up to normalization) since the Jacobian eigenvalues $\lambda^{(\eta)}_0$ are assumed to be non-degenerate.
In particular, when \( \Lambda^{(1)} = 0 \) the Jacobian eigenvector becomes  \( \boldsymbol{\phi}^{(1)}_0 = (\boldsymbol{1}^\top, -\boldsymbol{0}^\top)^\top \).

\subsubsection{Case 2: Eigenvalues \( \lambda^{(\bar{\alpha})}_0 = -\gamma + \tilde{\beta} S^* + D_I \Lambda^{(\alpha)} \)}

Substituting \( \lambda^{(\bar{\alpha})}_0 = -\gamma + \tilde{\beta} S^* + D_I \Lambda^{(\alpha)} \) into the second row, we have:
\[
\left[D_I \boldsymbol{\mathcal{L}} - \left(\gamma - \tilde{\beta} S^*\right) \mathbb{I}_\Omega\right] \boldsymbol{\phi}_{I,0}^{(\bar{\alpha})} = \left(-\gamma + \tilde{\beta} S^* + D_I \Lambda^{(\alpha)}\right) \boldsymbol{\phi}_{I,0}^{(\bar{\alpha})}.
\]
If \( \boldsymbol{\phi}_{I,0}^{(\bar{\alpha})} \) is the zero vector, both sides of the equation vanish. Otherwise, if \( \boldsymbol{\phi}_{I,0}^{(\bar{\alpha})} \) is an eigenvector of \( \boldsymbol{\mathcal{L}} \) associated with the eigenvalue \( \Lambda^{(\alpha)} \), and the left-hand side becomes
\(
D_I \Lambda^{(\alpha)} \boldsymbol{\phi}_{I,0}^{(\bar{\alpha})} - (\gamma - \tilde{\beta} S^*) \boldsymbol{\phi}_{I,0}^{(\bar{\alpha})}.
\)
Thus, the second row equation holds, confirming that \( \boldsymbol{\phi}_{I,0}^{(\bar{\alpha})} \) is consistent with \( \lambda^{(\bar{\alpha})}_0 \).
Now, substituting \( \lambda^{(\bar{\alpha})}_0 = -\gamma + \tilde{\beta} S^* + D_I \Lambda^{(\alpha)} \) into the first row, and after rearranging terms we have:
\[
(\gamma - \tilde{\beta} S^*) \boldsymbol{\phi}_{I,0}^{(\bar{\alpha})} = (D_I - D_S) \Lambda^{(\alpha)} \boldsymbol{\phi}_{S,0}^{(\bar{\alpha})} - (\gamma - \tilde{\beta} S^*) \boldsymbol{\phi}_{S,0}^{(\bar{\alpha})}.
\]
Dividing through by \( \gamma - \tilde{\beta} S^* \) (recalling it is nonzero), we find:
\[
\boldsymbol{\phi}_{I,0}^{(\bar{\alpha})} = E_\alpha \boldsymbol{\phi}_{S,0}^{(\bar{\alpha})}.
\]
where we have defined 
\[ E_\alpha = \dfrac{(D_I - D_S) \Lambda^{(\alpha)} - (\gamma - \tilde{\beta} S^*)}{\gamma - \tilde{\beta} S^*}. \]
Therefore, we can conclude that for the eigenvalues \( \lambda^{(\bar{\alpha})}_0 = -\gamma + \tilde{\beta} S^* + D_I \Lambda^{(\alpha)} \), the corresponding eigenvectors of Jacobian \(\tilde{\mathbf{J}}\) are: 
\begin{equation}
    \boldsymbol{\phi}^{(\bar{\alpha})}_0 = \begin{pmatrix} \boldsymbol{\Phi}^{(\alpha)} \\[.1cm] E_\alpha \boldsymbol{\Phi}^{(\alpha)} \end{pmatrix}.
\label{eq:eigvec_2}
\end{equation}
As before, if we consider the solution \( \boldsymbol{\phi}_{I,0}^{(\bar{\alpha})} = \mathbf{0} \), then the corresponding equation implies \( \boldsymbol{\phi}_{S,0}^{(\bar{\alpha})} = \mathbf{0} \), which contradicts the fact that eigenvectors forming a basis must be nonzero. As a special case of interest, let us notice that when \( \Lambda^{(1)} = 0 \), the expression for \( E_\alpha \) simplifies to
\(
E_1 = -1.
\)
Thus, in this case, \( \boldsymbol{\phi}^{(\Omega+1)} = (\boldsymbol{1}^\top, -\boldsymbol{1}^\top)^\top \), indicating a specific structure of the Jacobian eigenvector when \( \Lambda^{(1)} = 0 \).

\subsection{Stability of the DBMF scenario}

The structure of the two classes of Jacobian eigenvalues reveals that the most favorable stability condition occurs when all eigenvalues are negative, except \( \lambda^{(1)}_0 \equiv \Lambda^{(1)} = 0 \), which necessarily arises due to the underlying system dynamics. The presence of this marginal eigenvalue complicates the analysis, as linear stability is inconclusive, thereby requiring advanced tools such as center manifold theory~\cite{guckenheimer2013nonlinear}. In the following, we show that the complementarity of the two species, \( S \) and \( I \), naturally eliminates the need for such technical considerations.

We begin by considering the general solution for the evolution of perturbations:
\[
\begin{pmatrix}
  \delta \textbf{S}  \\
  \delta \textbf{I}
\end{pmatrix} = \sum_{\eta=1}^{2\Omega} C_{\eta} e^{\lambda^{(\eta)}_0 t} \boldsymbol{\phi}^{(\eta)}_0.
\]
Notably, due to the conservation relationship at the level of community nodes
\(
\sum_{\mu} S_{\mu} + I_{\mu} = N,
\)
it follows that
\(
\delta S_\mu = -\delta I_\mu,\,\forall \mu,
\)
meaning that within a single metanode, any decrease in \( S \) must correspond to an equivalent increase in \( I \), and vice versa. Consequently, the perturbations at \( t = 0 \) can be expressed in terms of eigenvectors as
\[
\begin{pmatrix}
  -\delta \textbf{I}  \\
  \delta \textbf{I}
\end{pmatrix} = \sum_{\eta=1}^{2\Omega} C_{\eta} \boldsymbol{\phi}^{(\eta)}_0.
\]
Expanding the Jacobian eigenvectors \( \boldsymbol{\phi}^{(\alpha)}_0 \) according to their structure, we obtain:
\begin{equation*}
\begin{aligned}
    \begin{pmatrix}
  -\delta \textbf{I}  \\
  \delta \textbf{I}
\end{pmatrix} &= C_1 \begin{pmatrix} \mathbf{1} \\ \mathbf{0} \end{pmatrix} 
+ \sum_{\alpha=2}^{\Omega} C_\alpha \begin{pmatrix} \boldsymbol{\Phi}^{(\alpha)} \\ \mathbf{0} \end{pmatrix} 
+ C_{\Omega+1} \begin{pmatrix} \mathbf{1} \\ -\mathbf{1} \end{pmatrix}\\ 
&+ \sum_{\alpha=2}^{\Omega} C_{\Omega + \alpha} \begin{pmatrix} \boldsymbol{\Phi}^{(\alpha)} \\ E_\alpha \boldsymbol{\Phi}^{(\alpha)} \end{pmatrix}.
\end{aligned}
\end{equation*}

Separating the components \( -\delta \textbf{I} \) and \( \delta \textbf{I} \), we have:

\begin{align*}
&\textit{First row:}\quad -\delta \textbf{I} = C_1 \mathbf{1} + \sum_{\alpha=2}^{\Omega} \left(C_\alpha + C_{\Omega + \alpha}\right) \boldsymbol{\Phi}^{(\alpha)} + C_{\Omega+1} \mathbf{1}, \\
&\textit{Second row:}  \quad \delta \textbf{I} = -C_{\Omega+1} \mathbf{1} + \sum_{\alpha=2}^{\Omega} C_{\Omega + \alpha} E_\alpha \boldsymbol{\Phi}^{(\alpha)}.
\end{align*}
After equating and simplifying, we obtain:
\[
C_1 \mathbf{1} + \sum_{\alpha=2}^{\Omega} \left(C_\alpha + C_{\Omega + \alpha}\right) \boldsymbol{\Phi}^{(\alpha)}
= -\sum_{\alpha=2}^{\Omega} C_{\Omega + \alpha} E_\alpha \boldsymbol{\Phi}^{(\alpha)}.
\]
To satisfy this equation for all \( \boldsymbol{\Phi}^{(\alpha)} \), it follows that  
\[
C_1 = 0.
\]
Recalling that \( E_1 = -1 \), we obtain the general relationship among the integration constants:  
\begin{equation}
C_\alpha = -(1 + E_\alpha) C_{\Omega + \alpha}, 
\label{eq:consts}
\end{equation}
for \( \alpha = 1, \dots, \Omega \).

In conclusion, in the DBMF framework, the state evolution of perturbations is given by:
\[
\lim_{t \to \infty} \left[\begin{pmatrix}
  \delta \textbf{S}  \\
  \delta \textbf{I}
\end{pmatrix} = \sum_{\eta > 1} C_\eta e^{\lambda^{(\eta)}_0 t} \boldsymbol{\phi}^{(\eta)}_0\right]= 0.
\]
since as \( t \to \infty \), the terms \( e^{\lambda^{(\eta)}_0 t} \to 0 \) for \( \eta > 1 \), implying that the perturbations \( \delta S_\mu \) and \( \delta I_\mu \) decay to zero, confirming that the system in this scenario is asymptotically stable.


\section{Numerical validation of the localization reduction}
\label{sec:local_valid}

\begin{figure*}
    \centering
    \includegraphics[width=\textwidth]{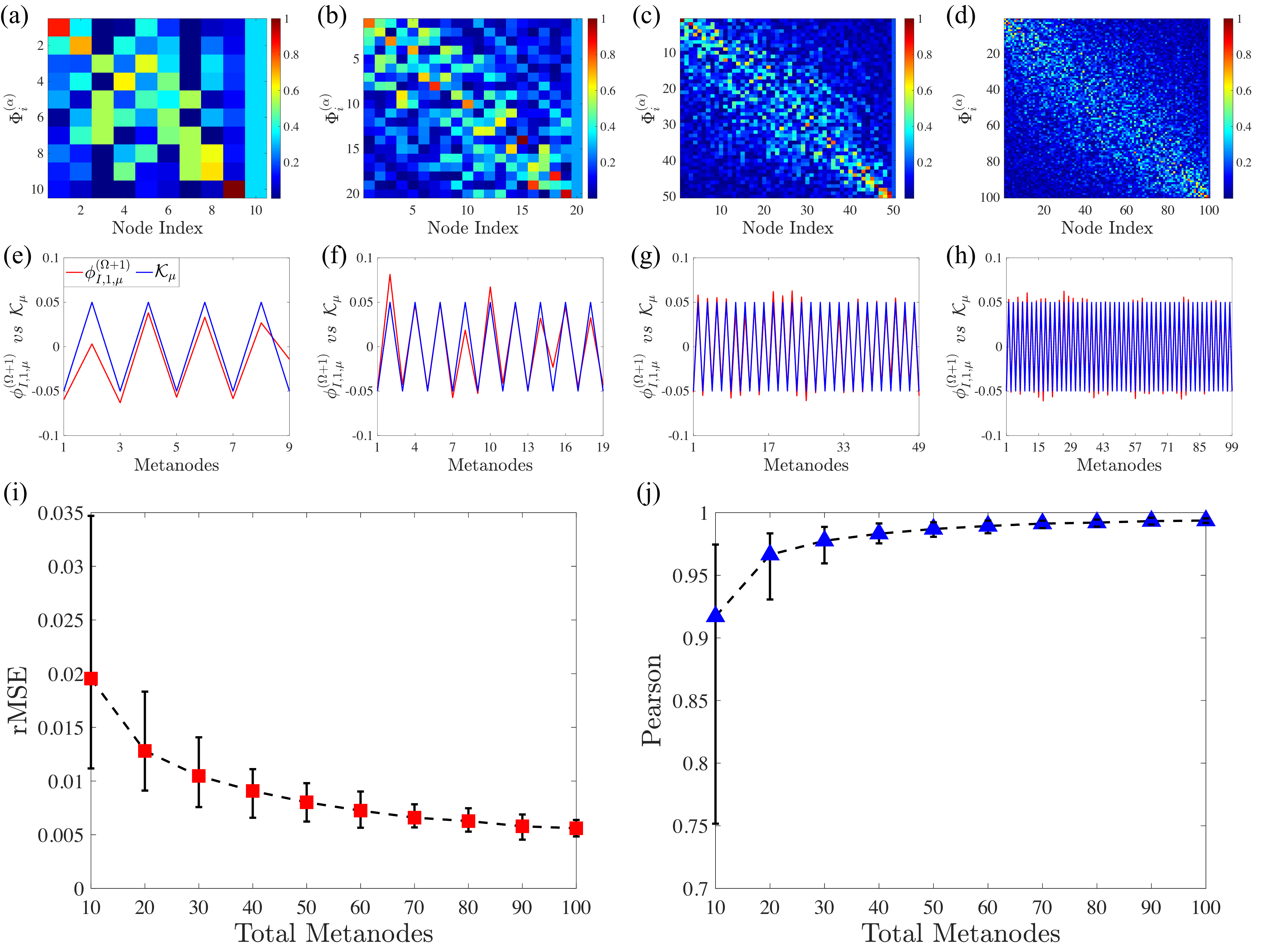}
    \caption{Localization and structural matching of Laplacian eigenvectors.
Panels~(a)--(d) show heatmaps of the absolute values of the Laplacian eigenvector components for networks with 10, 20, 50, and 100 nodes, respectively. Each column represents an eigenvector, ordered by increasing eigenvalue. Panels~(e)--(h) display the entry-wise comparison between the eigenvector \( \boldsymbol{\phi}^{(\Omega+1)}_{I,1} \) and the structural vector \( \boldsymbol{\mathcal{K}} \), generated as a sawtooth function, for each corresponding network size. Panels~(i) and~(j) report the root-mean-square error (rMSE) and the Pearson correlation, respectively, between \( \boldsymbol{\phi}^{(\Omega+1)}_{I,1} \) and \( \boldsymbol{\mathcal{K}} \), averaged over 50 independent network realizations with fixed parameters. Error bars indicate standard deviations across realizations.}
    \label{fig:Fig5}
\end{figure*}

To numerically validate the reduction results shown in Fig.~\ref{fig:Fig5}, panels~(e)--(h), we compared the eigenvector \( \boldsymbol{\phi}^{(\Omega+1)}_{I,1} \) with the structural vector \( \boldsymbol{\mathcal{K}} \), both derived and introduced in the main text. The vector \( \boldsymbol{\mathcal{K}} \) was generated here as a sawtooth function to avoid case-specific bias among different network realizations, unlike the data-driven construction used in the main paper. The comparison, based on Eq.~(8) of the main text, reveals an increasingly strong alignment as the network size grows. This trend is also confirmed quantitatively in panels~(i) and~(j), which report the root-mean-square error (rMSE) and the Pearson correlation coefficient, respectively. The latter is computed by centering both vectors and evaluating their normalized covariance, using the standard definitions~\cite{james2013introduction}:
\begin{align} 
\text{rMSE} = \sqrt{ \frac{1}{N} \sum_{j=1}^N \left( \phi_j - \mathcal{K}_j \right)^2 }, \\
\rho = \frac{ \sum_{j=1}^N (\phi_j - \bar{\phi})(\mathcal{K}_j - \bar{\mathcal{K}}) }{ \sqrt{ \sum_{j=1}^N (\phi_j - \bar{\phi})^2 } \sqrt{ \sum_{j=1}^N (\mathcal{K}_j - \bar{\mathcal{K}})^2 } }.
\end{align}

The results over 50 independent network realizations confirm the same trend: larger networks exhibit better agreement between \( \boldsymbol{\phi}^{(\Omega+1)}_{I,1} \) and \( \boldsymbol{\mathcal{K}} \).

\bibliographystyle{apsrev4-2}
\bibliography{draft_paper}

\end{document}